# Software Engineering for Internet of Things: The Practitioners' Perspective

Mahdi Fahmideh, Aakash Ahmad, Ali Behnaz, John Grundy, Willy Susilo

**Abstract—** Internet of Things based systems (IoT systems for short) are becoming increasingly popular across different industrial domains and their development is rapidly increasing to provide value-added services to end-users and citizens. Little research to date uncovers the core development process lifecycle needed for IoT systems, and thus software engineers find themselves unprepared and unfamiliar with this new genre of system development. To ameliorate this gap, we conducted a mixed quantitative and qualitative research study where we derived a conceptual process framework from the extant literature on IoT, through which 27 key tasks for incorporation into the development processes of IoT systems were identified. The framework was then validated by means of a survey of 127 IoT practitioners from 35 countries across 6 continents with 15 different industry backgrounds. Our research provides an understanding of the most important development process tasks and informs both software engineering practitioners and researchers of the challenges and recommendations related to the development of next-generation of IoT systems.

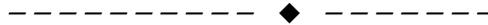

## 1 INTRODUCTION

According to its proponents, Internet of Things (IoT) is becoming a major phenomenon in recent years. Smart cities, smart grids, wearable devices, smart agriculture, Industry 4.0, and self-driving cars are some examples of application domains that IoT promises to revolutionize [1],[2],[3]. An IoT system potentially includes thousands of distributed smart objects, backbone services, platforms, and applications that are interconnected to provide added-value and intelligent reasoning for other systems and to end-users [4]. Design and implementation of this class of systems have now become mainstream in both academia and for smart cities, retail, logistics, manufacturing, agriculture, healthcare, and so on [5]. This momentum is concomitant to an ever-increasing number of dedicated international venues, such as journal special issues, conference tracks, and workshops, for software engineering of IoT, that has stimulated the motivation for our current research. Software development processes-, e.g. process frameworks or engineering methodologies-, are an integral part of the software engineering discipline and the centerpiece of quality management initiatives to develop and maintain software systems in a cost-effective manner [6], [7]. As the software engineering discipline is periodically faced with major new emerging technologies such as IoT [8], it is necessary to continually rethink how to develop systems in response to a shift to new paradigms and if conventional ways of software engineering are still applicable.

Initially, IoT systems were not viewed as a new concept, i.e., a world of connected things that can share data, but as theoretically similar to an end-to-end architecture. One might thus expect that the development process of IoT systems would have drawn on the existing software engineering body of knowledge, which has been accumulated, matured, applied, and tested thoroughly over time. There are occasional references confirming this viewpoint and belief that the IoT system development as so-called *old petunias in new bowls*, e.g., [9], [10],[11],[12]. However, there is a lack of empirical support for this argument. At the flip side, opponents, e.g.,[13],[14], [15], believe that- in contrast to the conventional software engineering which takes place predominantly within an individual system- the socio-technical nature of IoT systems raises a range of new important complexities. For example, various hardware and software components, each with their own distinct development process, are linked to different parties and are working in an extremely dynamic and distributed fashion. They incorporate requirements from various stakeholders such as citizens, legislators, and administrative authorities. Whilst some of these complexities are rooted in the immaturity of current technologies, some are intrinsic to IoT. The development of IoT systems is thus much more complicated than that typically deemed in the conventional software development wisdom [4],[5]. Confronted with the recent and rapid development of various IoT systems in different

- *Mahdi Fahmideh is with the School of Computing and Information Technology at University of Wollongong, Australia, e-mail: mahdi@uow.edu.au*
- *Aakash Ahmad is with the Department of Computer and Information Science, College of Computer Science and Engineering, University of Ha'il, Ha'il, Saudi Arabia, email: a.abbasi@uoh.edu.sa*
- *Ali Behnaz is with the Citibank, Sydney, Australia, email: ali.behnaz@gmail.com*
- *John Grundy is with the Faculty of Information Technology, Monash University, Melbourne, Australia, e-mail: john.grundy@monash.edu*
- *Willy Susilo is with the School of Computing and Information Technology at University of Wollongong, Australia, email: wsusilo@uow.edu.au*

domains, many software engineers have started to query how IoT systems should best be developed. For instance, Tecnalia et al. raise the issue that the majority of research done in this area has been focused on purely technical, i.e., implementation-oriented, aspects of IoT system development. A high-level process-oriented understanding and its related issues still remain few and far between [4]. Moreover, Fahmideh et al, point out that a holistic picture of the development process and different tasks to take into consideration for IoT system development is non-extant in the current software engineering literature [16].

As with any emerging software engineering paradigm, the successful development of IoT-based systems not only implies providing new tools and technologies, but also an understanding of the way of development cycle needs to aid software engineers to achieve reliable and systematic implementation of these systems [5]. This is confirmed by Giray et al. [14] who point out that *similar to the development of other systems, it is important for IoT systems to be developed in a systematic manner in order to achieve a proper system with respect to both the functional and non-functional requirements*. This needs to zoom out beyond various underlying technicalities around IoT and to identify an overarching view of the development process for this class of systems. Given the seeming absence of a collective and broad, yet IoT specific, development process view [5], our key research questions are:

RQ1. What tasks are suggested in the literature about IoT system engineering for incorporation into the development process of IoT systems?

RQ2. How do IoT experts perceive the importance of these development process tasks?

RQ3. How do IoT experts believe these tasks are important and become challenging in practice?

RQ4. What overall recommendations do IoT experts suggest to conduct these tasks?

To address the above research questions, we adopted a mixed qualitative and quantitative research method organized in two-phase of exploratory and confirmatory. In the first phase, we drew from the extant material in IoT literature a generic platform-agnostic process framework of phases and tasks. In the second phase, we examined the validity of the resultant probing framework and its soundness via conducting a Web-based survey of randomly selected experienced IoT software engineering practitioners. We engaged a total of 127 practitioners - based in 35 countries across 06 continents - with diverse professional backgrounds, including IoT architects, developers, project managers, with experience across diverse IoT sectors such as robotics, home automation, and transportation. Both phases of this research make significant contributions to the limited literature on software engineering of IoT systems. As further discussed later during the literature review, previous studies have not had the opportunity to explore and identify a development process framework attuned with IoT and did not offer such an empirical phase. No prior research has been undertaken to explicitly focus on the aspect of the software engineering development process for IoT systems. As will be discussed in Section 6, some researches [17], [18], [19] discuss enabling Internet-based computing technologies, such as cloud computing and data analytics, that provide backbone services to implement IoT systems. The second stream of studies, including [20], [21], [22], [23], presents an initial effort to help software engineers to better understand IoT system development. However, they are very broad and lack an empirical evaluation by experts and fail to highlight key development process tasks, as presented in this research. Hence, we deem our work as the first **comprehensive** effort that explicitly addresses and defines an IOT-based software engineering process, gains empirical feedback on this process, and has significant implications for the development of IoT systems in real-world-scenarios. This paper provides a description of how IoT development processes are understood and their relevancy for software teams, which makes it a pioneer revelatory research in its kind in turn. Our primary contributions are:

• Introducing an *empirically derived conceptual process framework,* including 27 common tasks organized into 3 phases along with their definitions, manifested themselves into the development processes of IoT systems;

• Providing *quantitative justifications of how* the framework's tasks are perceived important by IoT experts along with *contextual example* highlighting as to why these tasks are important and problematic during the development process of IoT systems;

• Sharing the *overall recommendations* for software engineering individuals and teams to consider in real-world IoT system development scenarios; and

• Constructing potential *research directions* for the software engineering research community to build supports for IoT developments and address emerging and futuristic challenges of IoT systems.

This paper brings both research and practical implications. Our framework can be viewed as an appraisal instrument enabling software teams to examine the extent to which their in-house approach supports IoT system development phases and tasks and, if necessary, apply augmentation. Moreover, those who are interested to join software engineering for IoT or in training and need reliable guidance can use the framework. Furthermore, the framework can be used as a starting point to study the engineering process of IoT systems regardless of their underlying enabling technologies and application domains.

In Section 2 we explain the research method used following by the presentation of the framework supplemented by the findings from domain experts in Section 3. Section 4 discusses the implications of this research. The threats, limitations, and further works of this research are presented in sections 5 and 6. We summarized key learnings in Section 7.

## 2  RESEARCH METHOD

To answer the research questions above, we use both qualitative and quantitative research methods, i.e., a mixed methods. As some prior researches acknowledged, e.g. [24], the mixed methods is recognized to be useful to get breadth and depth of understanding and corroboration of a phenomenon for which extant research is fragmented, premature, or equivocal. Given IoT area is in the relative infancy stage, mixed methods research increases the reliability and accuracy of our research findings. Using this method was justified by our initial screening and our pilot review of literature, where we found a limited number of studies answering our stated research questions. The second reason was the lack of consistent and agreed terminologies in the field [5], which makes identifying development process tasks challenging. We followed Venkatesh's guidelines [25] in designing the mixed methods presented in Fig.1 and detailed in subsections 2.1 and 2.2.

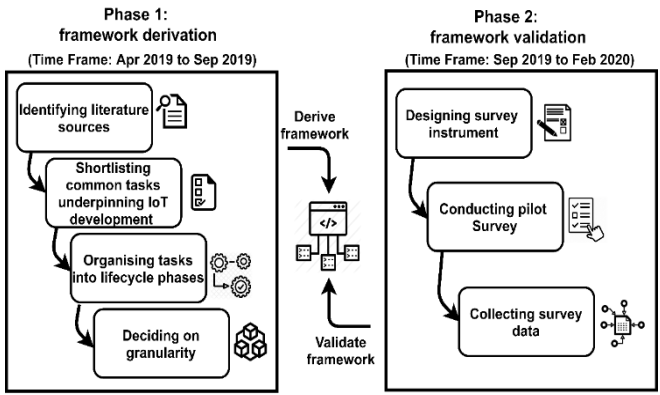

Fig.1. Research method

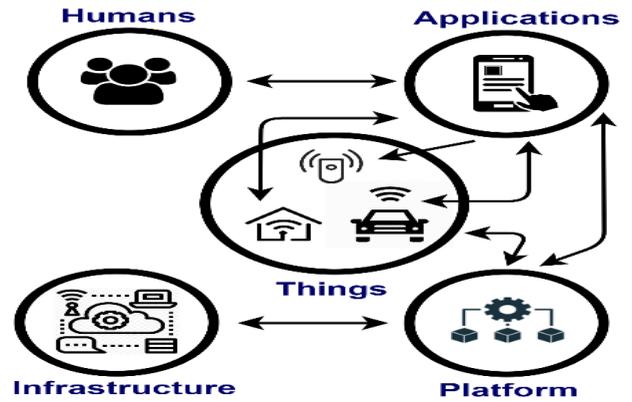

Fig .2. Fundamental constituents of IoT systems

## 2.1 Phase 1— framework derivation

### 2.1.1 IoT system conceptualization

Typically, new system development process approaches are grounded on some fundamental concepts and underlying logic. IoT systems are understood as a vision of a more connected world. For example, Cisco coined the term IoT as *the network of physical objects that contain embedded technology to communicate and sense or interact with their internal states or the external environment* [26]. An agreed definition of IoT, offered by the joint technical committee of the International Organization for Standardization (ISO) and the International Electrotechnical Commission (IEC), is *an infrastructure of interconnected objects, people, systems and information resources together with intelligent services to allow them to process information about the physical and the virtual world and react* [27]. These definitions are commonly rooted in a few key elements that an IoT system is built upon (Fig. 2) according to the literature [5],[15],[28]. The term *things (or objects)* refers to small/large physical trackable, configurable, and controllable objects such as sensors, smart objects, devices, humans, RFID tags, and buildings, with the ability to store, process, transmit or receive data. *Software applications*, such as web portal, mobile app or legacy systems, enable users to use functionalities provided by IoT system. An IoT platform including middleware, backbone services, and Application Programming Interfaces (APIs) give end-users access working with the IoT system for data collection, storing, analysis, and visualization. Another key fundamental element of an IoT system is its *infrastructure* which provides hardware for data processing, storing, computing, and interconnectivity among data centers, servers, and things via different network communication protocols. There are other elements such as *people* who connect to the system and use its provided functionalities. Collectively viewed, applications, things, platforms, and infrastructure are a combination of software and hardware components responsible to manage context-aware interactions with people. These fundamental elements are the basis for our proposed framework.

### 2.1.2 Steps for framework derivation

Our framework derivation was in line with RQ1 and performed between April 2019 and September 2019. To identify a generic process model, we initially started with reviewing existing literature survey papers that could lead us to continuous engagement and hermeneutic conceptual understanding of IoT development process [29]. This could help us avoid missing related important studies and forming a foundation for identifying relevant studies. Next, we conducted the following steps.

**Step 1. identifying literature sources**. We used guidelines for conducting the systematic literature review [30] as a point of departure to identify relevant studies to RQ1 followed with strict adherence to PRISMA (Preferred Reporting Items for Systematic reviews and Meta-Analyses) [31] as depicted in Fig.3 and elaborated as follows.

*Eligibility criteria.* Inclusion and exclusion criteria for study selection were defined. A paper could be selected if it:

(i) was published between 2010 and 2019. To manage the complexity of literature search in the voluminous IoT literature, we set this interval to reduce the number of papers to be identified in the first iteration and then later we used the *snowballing technique* [32] to identify papers earlier than 2010;

(ii) was published in software engineering or information systems related journals/conferences proceedings or by leading IoT solution providers such as Oracle, IBM, and Amazon.

(iii) described explicitly its research context, goals, methods, and results;

(iv) focused, either partially or fully, on the IoT-based software development process

We excluded the papers that:

(i) were in non-English language; and

(ii) were too general or short introductory papers that couldn't provide input for the framework derivation steps.

*Data sources.* The common scientific digital libraries IEEE Explore, ACM Digital Library, SpringerLink, ScienceDirect, Wiley InterScience, ISI Web of Knowledge, and Google Scholar were set as sources for the literature search. They contain by far the majority of publications on IoT and software engineering approaches. Furthermore, we took into account international venues in leading information systems and software engineering journals and conference proceedings mainly IEEE Transactions on Software, Engineering/Cloud Computing (TSE and TCC), JSS, IST, TOSEM, the senior basket of eight IS journals, SIG recommended journals by associated of information systems (AIS) (See Appendix A). Regarding RQ1, we paid particular attention to the proceedings of international conferences, symposiums, and workshops dedicated to the IoT domain. In particular, we sought papers published in IEEE IoT (IoT-J) and recently established venue of ACM transaction of IoT (TIOT). As recommended by Ogawa et al. [33], we did not neglect the importance of grey literature such as internet blogs, white papers, and trade journal articles, that could lead us to insights on the IoT software engineering development process as listed in Appendix A. We used an abbreviation for each study, based on its proposal whether it was a model, approach, platform, etc., along with the identifier S and a number, denoted as [S#], to refer to the studies that we identified and used as input for further steps.

*Search strings.* Initially, we defined combined search strings based on *IoT* AND (*approach, method, methodology, system development method, process, development process, process model, process lifecycle,* and *framework*). However, our initial screening of the literature showed that the information on the development process of IoT systems was scattered over different themes. For example, we discerned that a fragment of IoT literature knowledge is attuned with IoT platform development [5] where papers related to this theme could present different functionalities to be realized by an IoT system and hence imply specific development tasks. We added the term *platform* to our updated search strings.

*Study selection.* As shown in Fig. 4. a total of 250 papers were found by our initial search in both academic and grey literature. Following with forward and backward search recursively for each identified paper's references via snowballing [32], 53 new papers were added to our study list. This resulted in 303 papers that were reduced to 295 after removing 8 duplicated ones from the same authors. We screened the title, abstracts, and in most cases, the full texts of the retrieved papers, and this resulted in 169 papers based on the eligibility criteria. We came up with 66 papers, 17 from the grey literature, and 49 from the academic literature, which met the inclusion criteria for our framework derivation defined above. A summary of the identified papers' demographic information is shown in Figure 4 but the full details are available in Appendix A.

**Step 2. shortlisting common tasks underpinning IoT development**. The identification of tasks was an iterative and gradual process. In the first iteration of this step, we reviewed all 66 identified papers from the previous step where, for each one, we tried to identify text segments that could be explicitly or implicitly labelled as an individual task for the inclusion into the target process framework. We set three criteria for extracting a text segment from a paper as a potential task:

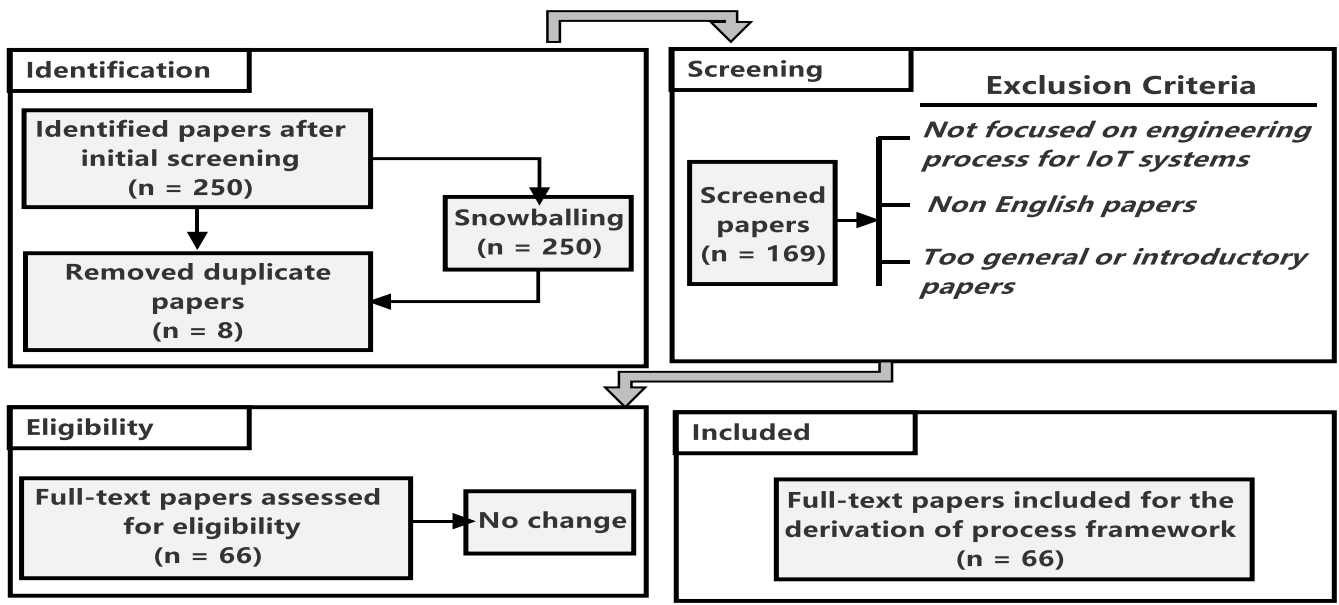

Fig. 3. PRISMA flow through the different steps of a systematic review

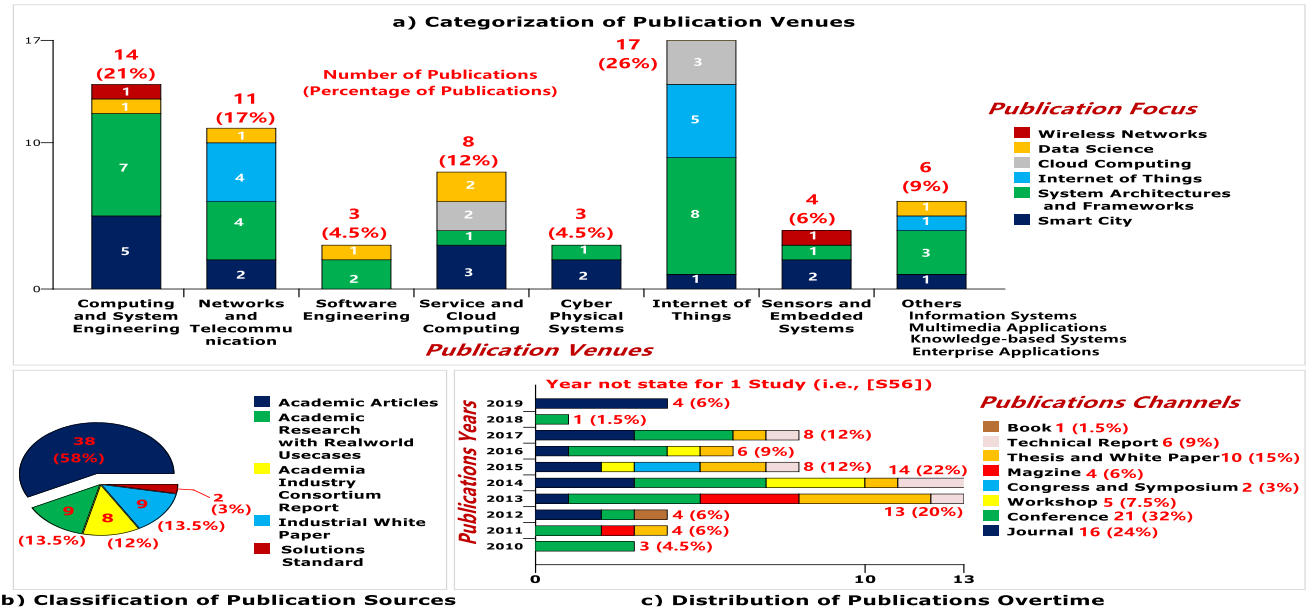

Fig. 4. Demographic information of identified studies

(i) Being sufficiently generic to a variety of IoT system domains regardless of underlying technologies;

(ii) Widespread recurring themes in the paper set as an indicator of its importance for inclusion in the framework

(iii) Traced to the fundamental elements of IoT systems as discussed in Section 2.1.1.

We discarded tasks related to the maintenance phase as this was out of the scope of our current proposed framework. We were not interested in identifying roles and work-products that could be associated to the tasks. We excluded the tasks that were related to governance, such as risk management, quality assurance, configuration management, and measurement. These all were out of the scope of our current research and considered as future work as discussed in Section 4.2.

According to the first criterion (i), problem-specific or technology-oriented tasks were omitted. For instance, using *Apache CouchDB* or *Hadoop distributed file system* design in CiDAP [S3] (Appendix A) discusses operationalization data management mechanisms in the design phase. We found that not all IoT systems may necessarily adopt these technologies and, hence, they were deemed too specific for inclusion in the framework and thus they were not extracted from CiDAP [S3]. With respect to the second criterion (ii), we took into account the frequent patterns of any particular text segment, theme, or definition in a paper as a yardstick to decide to code that text fragment as a task grounded in what other papers commonly referred to as. As an illustration, the task *Resource discovery design* in the resultant framework, which concerns with providing publish/subscribe mechanisms in system architecture to enable discovering and to allow smart objects to connect the IoT system, is jointly pointed out by studies VITAL [S2], CiDAP [S3], IoT-ARM [S8], OpenIoT [S9], FIWARE [S12], Telco USN-Platform [S18], GAMBAS [S25], Nitti [S43]. For criterion (iii), we checked if a text segment could be related to IoT fundamental concepts. The output of this iteration was an initial list of 384 labelled tasks including their original reference papers. They were stored in a spreadsheet [34]. This enabled us to have the track of task origination from the identified papers.

The second iteration aimed at synthesizing 384 fine grained tasks extracted from the first iteration to a higher-order set of distinct IoT software development process tasks. This was undertaken on the basis of inherent similarities of tasks and the fact that people working on IoT coming from different domains may use different terminologies and phrases to refer to identical things. For example, from studies VITAL [S2], IoT-ARM [S8], FIWARE [S12], Vilajosana [S13], Giang [S21], RERUM [S32], BASIS [S44], SORASC [S46], and IoTEP [S63], we observed that a task like *Interaction design*, collectively, deals with identifying integration points, data flow between hardware and software components, and how data is entered to the system and sent out. This step resulted in 27 tasks. These tasks are *Ideation*, *Domain requirement analysis*, *Infrastructure requirement analysis*, *Application requirements analysis*, 5. *Smart object requirement analysis*, 6. *Stakeholder analysis*, 7. *Plan definition*, 8. *Resource discovery design*, 9. *Data collection design*, 10. *Data cleaning design*, 11. *Data storing design*, 12. *Data processing design*, 13. *Query processing design*, 14. *Meta-data generation design*, 15. *Data visualization design*, 16. *Monitoring design*, 17. *Service composition design*, 18. *Event processing design*, 19. *Platform architecture design*, 20. *Smart object architecture design*, 21. *Interaction design*, 22. *Application architecture design*, 23. *Application coding*, 24. *Smart object coding*, 25. *Platform coding*, 26. *Testing*, and 27. *Installation*. Note that, tasks *Testing* and *Installation* included testing subtasks for applications, smart objects, and platforms but we summarized under one task name.

**Organising tasks into lifecycle phases**. A simple yet sufficiently generic core process framework which could encompass the identified IoT software engineering process tasks was targeted. Influenced by the generic software development lifecycle (SDLC) introduced by Pressman [35] that provides adequate high-level coverage on any classes of software system development, we set a simple lifecycle covering the phases of Analysis, Design, Implementation and Test, where each phase was populated by the identified tasks. Adopting such a generic SDLC in a top-down fashion for process framework design is consistent with prior researches, such as work in designing process metamodels for cloud computing migration [17], and the approach proposed for software language engineering in the agent-oriented software engineering research domain [36]. Placing the tasks in each phase was mainly based on their similarities and semantic relationships to a phase. For instance, the tasks related to *requirements analysis* and *plan definition* were positioned in *design phase* with an aggregation relationship, i.e., ━━◆ between the phases and tasks. The sequence between the phases is shown using ━━▶ symbol.

**Deciding on granularity**. In the development of any conceptual models, researchers face the issue of determining trade-offs between pros and cons of simple vs. complex model, i.e., too generic or too specific [17] ,[37]. Given that the priority of our framework was understandability over completeness and being adequately versatile, cohesive, and including distinct tasks with minimum overlapping, we did not include complex and domain-specific tasks in the framework to make it more generic and less detailed. We did not commit to narrow it down into operationalization of tasks via supportive techniques or modelling tools. Given the sheer volume of research suggesting techniques for IoT system development, as reviewed in Section 6.2, we deemed such a dedication would be premature. Moreover, we recognized it is intuitive that a task could be associated with one or more fundamental elements of an IoT system. For example, the task *coding* in the framework was founded as a coarse-grain overarching task as it is performed for elements software applications, smart objects, and platforms. We decided to define the *coding* as the super-class task in the framework where making the tasks *application coding*, *thing coding*, and *platform coding* as subclass tasks. This relationship was indicated via the specialization symbol, i.e., ━━▷, in the framework.

## 2.2 Phase 2— framework validation

To answer RQ2, RQ3, RQ4, our second phase was confirmatory where we conducted a survey to validate and to assess how the framework could be generalized to a broader population in the view of the research questions. Following the guidelines for survey design in Pinsonneault et. al, [38], we performed three steps as outlined below.

**Designing survey instrument**. The primary purpose of the survey was to examine if the framework's tasks are perceived as important by IoT experts. We used the Qualtrics [39] online survey software creator to design and conduct our survey. We provided the framework tasks along with their definitions. For RQ2, we asked respondents to rate the importance of each task based on seven Likert-scale from 1 to 7, where 1 and 7, respectively, indicated not at all important and extremely important. Regarding RQ3 and RQ4, we also provided *open-ended questions* asking respondents to give their reasons why these tasks are important, the most important IoT-specific challenges that software engineers may face during development compared to non-IoT system development, and, if any, recommendations. The respondents were also asked to suggest any missing important tasks that should be added to the framework. We also obtained the survey questions, consent form, participant selection technique, and survey data management which were reviewed and approved by our academic institute ethics authority.

**Conducting pilot survey**. We piloted our survey to check if it was as coherent and concise as possible whilst gathering sufficient feedback. Minor issues reported and addressed were related to spellings, IoT-specific terminologies, the sequence of questions and activities, and adding one question to the survey. The link to the survey is available at [40].

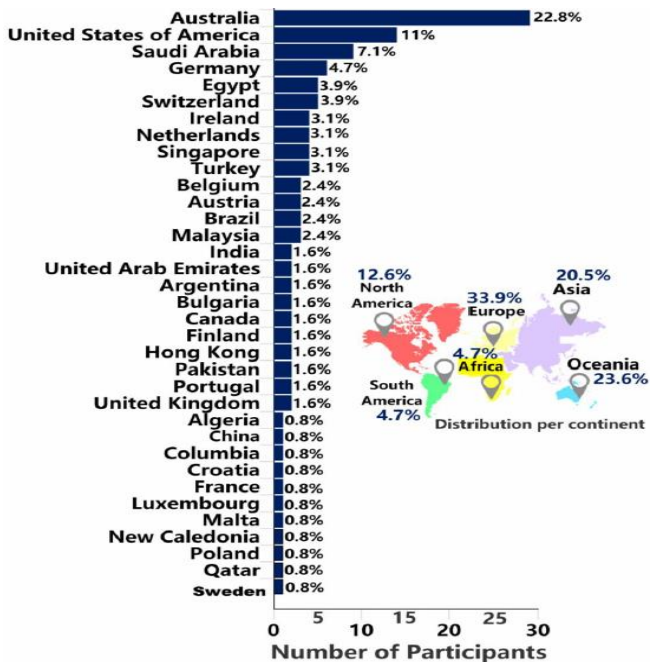

Fig. 5. Geographical distribution of participants

**Collecting survey data**. The survey was made online between September 2019 and February 2020 [40]. Since IoT is a nascent field, the entire population of IoT domain experts was unknown to draw a sample size for the data collection and it was difficult to identify qualified people with real-world experience in the development of IoT systems who could take part in the survey. Hence, we performed purposeful sampling [41] and manually screened the professional profiles across the world and purposefully selecting respondents who had real-world experience and had been involved in an end-to-end development process of, at least, one IoT system. We sought potential respondents in different online IoT communities, in particular, LinkedIn, Facebook, GitHub, and academic research groups. For example, we used GitHub APIs to mine commit logs in Github repositories to identify the email addresses of active IoT developers who had submitted changes and contributed to at least one IoT project. After checking her/his online pro-file, we confirmed the credibility of each selected respondent by directly asking she/he to provide their professional and demographic information, in particular, their roles, development process, and technologies they used during their IoT project. This allowed us to check the differences between opinions from various backgrounds. Once willingness and expertise were confirmed, an invitation letter along with the survey link was formally issued. In total, we sent this voluntary survey to a list of 340 randomly identified and verified IoT experts through personalized emails or direct messages.

Finally, 127 respondents provided complete answers with an overall response rate of 37%. In terms of the geographical distribution of the respondents from 35 different countries, an ascendant ordering of participants based on seven continents revealed the response rates for Europe with 33.9%, Oceania with 23.6%, Asia with 20.5%, North America with 12.6%, South America with 4.7%, and Africa with 4.7% as shown in Fig. 5. The majority of the participants were from Australia, the United States, and Saudi Arabia with a response rate of 22.8%, 11%, and 7.1%, respectively. We observed 3.1% response rate each from Ireland, Singapore, and Netherland.

We characterized the respondents with respect to their roles in IoT projects classified as the software programmer, software architect, hardware designer, software tester, project manager, consultant, security engineer, researchers, supporter, trainer, and industrial engineer. The majority of respondents, i.e., 69 and 46 out of 127, mentioned that they have been involved in IoT development processes as a software architect and software programmer, respectively. Additionally, the respondents were allowed to select multiple industry sectors, wherever relevant. We found that the respondents come from 14 industry sectors, among them, their experience was more related to IoT system development for logistics and transportation, agriculture and smart farming, manufacturing, smart energy, and utilities (Fig 6. a). The respondents were with an average of 5 years' of experience, accounting for a collective 656 years of experience in the field of IoT. Years of experience were classified into four groups, i.e., 45% between 1 and 3 years, 35% between 4 and 7 years of experience and. Finally, 2% and 20% of respondents had less than one and more than eight years of experience, respectively (Fig 6. b).

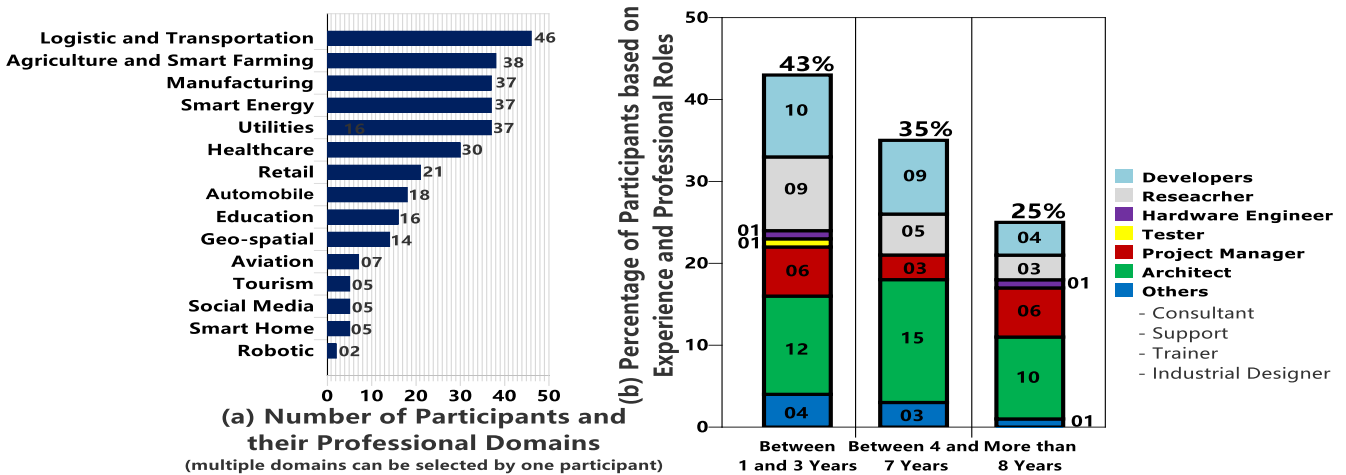

Fig. 6. a) Distribution of (a) participants' industry sectors and b) years of experience in IoT

# 3 FINDINGS

The following subsections 3.1, 3.2, 3.3, and 3.4 present the answers to the research questions RQ1, RQ2, RQ3, and RQ4, respectively, based on the findings obtained from the research phases as in Fig 1.

## 3.1 Process framework

We use an object diagram to represent our framework. In response to RQ1, Fig. 7 visualize the framework, including in total 27 tasks (note that some fine-granular tasks presented in Fig. 8 and Fig. 9), rooted in the literature but without being bounded to situation-specific operationalization and technical details. Definitions of framework tasks are summarized in Appendix B and in Section 3.4.

## 3.2 Developers' perception about process tasks

**Quantitative data analysis**. To answer RQ2, the following steps were performed. We used SPSS software to examine the perceived importance of tasks across the demographic groups. For each individual task in the framework, the null and alternative hypothesizes were defined: *H0: the mean of the importance of the task is less than 5 vs. H1: the mean of the importance of the task is more than 5*. That is, if a development task is rated equal to or more than mid-point 5, i.e., somewhat important in Likert-scale, it is perceived as important. We used One-Sample T-Test to check if their importance ratings of tasks were distributed away from the median 5. This test was suitable for determining if the mean of an independent variable sample data is different from a specific value. We checked T-Test's assumptions first. This included the assumption of approximate normality of the sample responses, i.e. tasks as the dependent variables, which was checked using Kolmogorov–Smirnov test [33]. The *p-value* of Kolmogorov–Smirnov test was less than 0.05 indicating the data had positive skew, i.e., a violation from the normality assumption. Hence, we leveraged the central limit theorem [42], i.e., a sufficiently large sampling distribution, N=127 in this case, has an approximate normal distribution. We checked this argument visually via the normal Q-Q plots generated by Kolmogorov–Smirnov test confirming that all the data points were close to the diagonal line. This argument has also been verified by simulation results showing the parametric tests are not sensitive to the non-normality assumption in general [35], [36]. One-Sample T-Test also assumes that the dependent variables should be measured at a ratio or interval level. Using the Likert-scale to measure the importance of the tasks met this condition. Finally, the test assumes that the responses are collected independently. This assumption holds as none of the respondents were aware of the identity of other participants in this survey. Below, we provide survey results on the relative importance of our framework's tasks and the view about these tasks across different demographic groups.

**Perceived importance of task**. Table 1 shows the descriptive statistics and the results of One-Sample T-Test's given p < .05 for each task. The *p* values in Table 2 were divided by 2 as the hypothesis was formulated for a right-tailed test. From column six of this table, it can be observed that, despite the diversity of respondents, they statistically rated the majority of the framework's tasks, i.e. 25 out of 27, important (p < .05): these tasks are important for consideration during the development process. Of the framework tasks, *Data cleaning design* (t-test statistic of=0.522, p-value of=0.3) and *meta-data generation design* (t-test statistic of=0.071, p-value of =0.47) were not regarded by the respondents as significant compared to other tasks. All the tasks were rated quite high, with even the lowest, i.e., *meta-data generation design*, and received an average rating of 5.01. The highest rated task is *test* (mean of 6.23), which is performed in *implementation phase* and indicating the importance of testing. *Ideation* (mean of 6.18) and *domain requirements analysis* (mean of 5.99) follow in rating. These demonstrate the importance of understating if an IoT system is a viable solution for a given context to solve a problem as well as the significance of elicitation, analysis and validation of requirements in an IoT system. As discussed in Section 3.3, the relatively high rating of these tasks suggests that respondents are most concerned about the understanding the added value of an IoT system development, its requirements, and the successful addressing of these requirements via testing. This implies that IoT system requirements gathering and testing are more complex than conventional software development. In contrast, the *meta-data generation design* is the lowest rated task (Table 2). Available tools for automatic generation of different meta-data, making it transparent to developers, may help explain this low rating. Averages of rates for each phases of analysis, design and implementation are 5.74, 5.48 and 5.83. The respondents seem to believe implementation has more importance vs design.

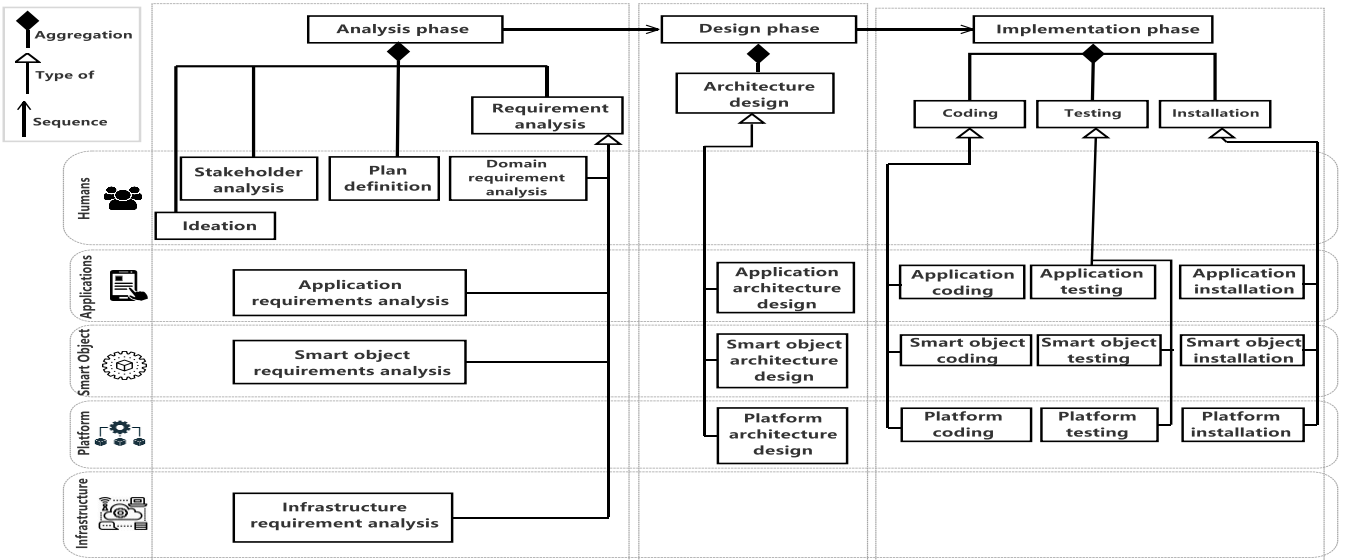

Fig. 7. Conceptual development process framework for developing IoT based systems

TABLE 1 Descriptive statistics for the framework tasks and the results of One Sample T-Test for each task

| Tasks | Mean | Std. deviation | T-Statistic value | p-value |
|---|---|---|---|---|
| **Analysis** | | | | |
| 1. Ideation | 6.18 | 0.97 | 13.59 | .00 |
| 2. Domain req. analysis | 5.99 | 0.93 | 11.91 | .00 |
| 3. Infrastr. req. analysis | 5.60 | 1.01 | 6.62 | .00 |
| 4. App. req. analysis | 5.61 | 1.11 | 6.21 | .00 |
| 5. Smart obj. req. analysis | 5.39 | 1.07 | 4.03 | .00 |
| 6. Stakeholder analysis | 5.81 | 1.10 | 8.28 | .00 |
| 7. Plan definition | 5.62 | 0.99 | 7.07 | .00 |
| **Design** | | | | |
| 8. Resource dis. design | 5.34 | 1.18 | 3.22 | .00 |
| 9. Data collection design | 5.84 | 1.02 | 9.24 | .00 |
| 10. Data cleaning design | 5.07 | 1.52 | 0.52 | .30 |
| 11. Data storing design | 5.44 | 1.16 | 4.26 | .00 |
| 12. Data process. design | 5.59 | 1.28 | 5.19 | .00 |
| 13. Query process. design | 5.45 | 1.17 | 4.31 | .00 |
| 14. Meta-data gen. design | 5.01 | 1.24 | 0.07 | .47 |
| 15. Data vis. design | 5.61 | 1.13 | 6.02 | .00 |
| 16. Monitoring design | 5.67 | 0.98 | 7.67 | .00 |
| 17. Service composition | 5.21 | 1.19 | 2.00 | .02 |
| 18. Event processing | 5.56 | 1.05 | 5.99 | .00 |
| 19. Platform arch. design | 5.51 | 1.20 | 4.80 | .00 |
| 20. Smart obj. arch. des. | 5.87 | 1.02 | 9.51 | .00 |
| 21. Interaction design | 5.26 | 1.19 | 2.44 | .00 |
| 22. App. arch. design | 5.72 | 1.16 | 6.91 | .00 |
| **Implementation** | | | | |
| 23. Application coding | 5.76 | 1.11 | 7.62 | .00 |
| 24. Smart object coding | 5.80 | 1.11 | 8.03 | .00 |
| 25. Platform coding | 5.69 | 1.08 | 7.22 | .00 |
| 26. Testing (all types) | 6.23 | 0.75 | 18.26 | .00 |
| 27. Installation (all types) | 5.68 | 1.09 | 6.95 | .00 |

**Power analysis.** A post-hoc analysis was performed via the G*power 3 software [43] to check the statistical power of our performed One-Sample T-Test. The analysis takes the parameters effect size, i.e., d, sample size, and $\alpha$ error probability. Cohen [44] defines three levels to conceptualize the power, i.e., small effect (d = 0.2), moderate effect (d = 0.5), and large effect (d = 0.8). In this research, the alpha is 0.05, effect size d = 0.2, and sample size 127, were sufficient enough to achieve the statistical power 0.7 (1-$\beta$ error probability).

**Demographic analysis.** We looked for any demographic differences in the perceived importance of the tasks. This was conducted in two steps. First, with univariate analysis, we ran One-Way ANOVA test to check if the ratings of each task, as a dependent variable, would be different by each of the three demographic independent variables, i.e., the years of experience, geographical location, and practitioners vs. academia. The results shown in Table 3 indicate that, except for task *ideation* highlighted with gray, as determined by (F (3,123) = 5.08, p = .01), there was not a statistically significant difference in the ratings based on the groups of the respondents' years of experience, i.e., y<=3, 4=<y<=7, and years>=8. In the second step, i.e. multivariate analysis, the test was re-run to check if there is any overlap among the three group of years of experience regardless of the variables geographical location, and practitioners vs. academia. This would allow us to understand the unique amount of the variance in the task rating that could be attributed to each independent variable. A Tukey post-hoc test in the multi-variate analysis, confirmed this difference and revealed that the importance of task *ideation* was significantly different between the respondents with the years of experience between 1 and 3 compared to those with more than 8 years of experience (p = .01). In other words, in both univariate and multi-variate analysis, the importance of the task *ideation* was perceived differently, indicating that the respondents who had more experience in IoT projects rated the task higher than those with less.

We wanted to know if the respondents' geographical location would have impacted the ratings. We divided the respondents into six continents, and performed One-Way ANOVA test for each individual task. The test identified the difference in the ratings with respect to the continent. In Table 2, each grey highlighted row shows the different perceived importance by the respondents based on their continents. For example, the post-hoc multi-variate analysis shows that Asian respondents (N=28) rated the task *domain requirement analysis*, relatively higher than all other groups. Similar observations are also true for the tasks *IoT application requirements analysis*, *resource discovery design*, *query processing design*, and *IoT application code*. We refrain from drawing a conclusion on such a difference and discuss it in the research threats section 4.2.

We compared the importance of the ratings between academic respondents and practitioner ones as the academia are typically possess extensive theoretical knowledge in a field of research in terms of knowing how and why the newly introduced technology works whereas practitioner who often use, promote, or disseminate technical inventions [45]. The practitioners (N=105), i.e., developers, architects, testers, project managers, consider certain tasks more important than the academia (N=22) do. We used Independent-Samples T-test to check this. Table 2 shows that the practitioners are more strongly in favor of tasks *ideation* T (125) = 4.34, p=0.00, *stakeholder analysis* (T (125) = 2.57, p=0.01), *event processing design* (T (125) = 2.1, p=0.03), and *installation* (T (125) = 4.34, p=0.05) compared to academics. This implies that IoT practitioners focus on requirements analysis, which is less important from the perspective of academics.

TABLE 2. Univariate analysis between the demographic variables (independent variables) and the perceived importance of tasks, i.e., task rating (dependent variables)

| Task | Years of experience | | | | | | | | Continents | | | | | | | | | | | | | | Practitioners vs. academia | | | | |
|---|---|---|---|---|---|---|---|---|---|---|---|---|---|---|---|---|---|---|---|---|---|---|---|---|---|---|---|---|
| | y<=3 (N=59) | | 4=<y<=7 (N=43) | | y>=8 (N=25) | | $F$ | $p$-value | Oceania (N=30) | | Europe (N=43) | | South America (N=6) | | North America (N=16) | | Asia (N=26) | | Africa (6) | | $F$ | $p$-value | Practitioners (N=105) | | Academia (N=22) | | $T$ | $p$ |
| | Mean | Std. deviation | Mean | Std. deviation | Mean | Std. deviation | | | Mean | Std. deviation | Mean | Std. deviation | Mean | Std. deviation | Mean | Std. deviation | Mean | Std. deviation | Mean | Std. deviation | | | Mean | Std. deviation | Mean | Std. deviation | | |
| 1. Ideation | 5.93 | 1.01 | 6.26 | 0.90 | 6.64 | 0.86 | 5.08 | 0.01 | 6.07 | 1.10 | 6.17 | 1.07 | 6.43 | 0.97 | 6.31 | 0.60 | 6.14 | 0.97 | 6.33 | 0.81 | 0.25 | 0.93 | 6.34 | 0.84 | 5.41 | 1.22 | 4.34 | 0.00 |
| 2. Dom. req. analysis | 5.95 | 1.04 | 6.05 | 0.75 | 6.00 | 1.00 | 0.13 | 0.87 | 6.00 | 0.92 | 5.98 | 0.75 | 5.57 | 0.97 | 6.06 | 0.85 | 6.29 | 0.76 | 5.00 | 2.09 | 2.30 | 0.04 | 6.01 | 0.92 | 5.91 | 1.01 | 0.45 | 0.65 |
| 3. Infrastructure req. analysis | 5.54 | 1.03 | 5.56 | 0.93 | 5.8 | 1.11 | 0.61 | 0.54 | 5.76 | 1.15 | 5.37 | 1.09 | 5.43 | 0.53 | 5.31 | 0.94 | 5.82 | 0.86 | 6.33 | 0.51 | 1.81 | 0.11 | 5.70 | 1.03 | 5.14 | 0.83 | 2.38 | 0.01 |
| 4. Application req. analysis | 5.49 | 1.05 | 5.70 | 1.05 | 5.76 | 1.33 | 0.69 | 0.50 | 5.76 | 1.02 | 5.17 | 1.09 | 5.43 | 1.51 | 5.81 | 0.83 | 5.96 | 1.17 | 6.00 | 0.89 | 2.35 | 0.04 | 5.63 | 1.09 | 5.55 | 1.22 | 0.31 | 0.75 |
| 5. Smart object req. analysis | 5.33 | 1.00 | 5.44 | 0.99 | 5.39 | 1.07 | 0.10 | 0.89 | 5.69 | 1.03 | 5.29 | 0.84 | 5.14 | 1.06 | 4.81 | 1.27 | 5.61 | 1.16 | 5.33 | 1.36 | 1.79 | 0.11 | 5.36 | 1.11 | 5.50 | 0.91 | -0.54 | 0.58 |
| 6. Stakeholder analysis | 5.75 | 1.26 | 5.72 | 0.90 | 6.12 | 0.97 | 1.23 | 0.29 | 6.03 | 1.21 | 5.59 | 1.16 | 6.29 | 0.75 | 5.75 | 0.93 | 5.82 | 1.05 | 5.83 | 1.16 | 0.84 | 0.52 | 5.92 | 1.08 | 5.27 | 1.03 | 2.57 | 0.01 |
| 7. Plan definition | 5.47 | 1.07 | 5.72 | 0.85 | 5.8 | 1.00 | 1.27 | 0.28 | 5.83 | 0.96 | 5.56 | 0.97 | 5.71 | 0.75 | 5.31 | 0.87 | 5.61 | 1.22 | 5.83 | 0.40 | 0.65 | 0.66 | 5.63 | 1.05 | 5.59 | 0.59 | 0.16 | 0.87 |
| 8. Resource discovery design | 5.41 | 1.27 | 5.28 | 1.05 | 5.28 | 1.20 | 0.18 | 0.83 | 5.90 | 1.08 | 5.10 | 1.11 | 5.29 | 0.75 | 5.00 | 1.67 | 5.46 | 0.74 | 4.67 | 1.86 | 2.48 | 0.03 | 5.38 | 1.25 | 5.14 | 0.77 | 0.88 | 0.38 |
| 9. Data collection design | 5.90 | 1.04 | 5.84 | 0.81 | 5.72 | 1.30 | 0.26 | 0.77 | 5.83 | 1.10 | 5.93 | 0.87 | 5.86 | 0.69 | 5.88 | 1.08 | 5.82 | 0.90 | 5.33 | 2.16 | 0.34 | 0.88 | 5.89 | 1.01 | 5.64 | 1.09 | 1.03 | 0.30 |
| 10. Data cleaning design | 5.12 | 1.54 | 4.98 | 1.53 | 5.12 | 1.53 | 0.12 | 0.88 | 5.34 | 1.63 | 5.39 | 1.30 | 4.29 | 1.70 | 4.69 | 1.35 | 4.71 | 1.67 | 5.17 | 1.72 | 1.45 | 0.21 | 5.11 | 1.54 | 4.86 | 1.45 | 0.69 | 0.48 |
| 11. Data storing design | 5.39 | 1.16 | 5.60 | 1.05 | 5.28 | 1.37 | 0.71 | 0.49 | 5.62 | 1.08 | 5.44 | 1.02 | 5.00 | 1.15 | 5.06 | 1.34 | 5.68 | 1.12 | 5.00 | 2.00 | 1.08 | 0.37 | 5.49 | 1.21 | 5.23 | 0.86 | 0.94 | 0.34 |
| 12. Data processing design | 5.64 | 1.33 | 5.58 | 1.11 | 5.48 | 1.44 | 0.14 | 0.86 | 5.66 | 1.42 | 5.68 | 1.10 | 4.71 | 1.49 | 5.13 | 1.02 | 5.89 | 1.10 | 5.50 | 2.34 | 1.48 | 0.20 | 5.54 | 1.33 | 5.82 | 1.00 | -0.91 | 0.36 |
| 13. Query processing design | 5.46 | 1.22 | 5.49 | 0.93 | 5.36 | 1.44 | 0.09 | 0.90 | 5.79 | 1.17 | 5.49 | 0.97 | 4.86 | 0.69 | 5.06 | 0.85 | 5.61 | 1.39 | 4.50 | 1.76 | 2.19 | 0.05 | 5.50 | 1.16 | 5.18 | 1.22 | 1.17 | 0.24 |
| 14. Meta-data gen. design | 4.98 | 1.30 | 4.77 | 1.15 | 5.48 | 1.15 | 2.68 | 0.07 | 5.34 | 1.17 | 5.02 | 1.06 | 4.86 | 1.06 | 4.50 | 1.41 | 5.00 | 1.33 | 4.83 | 1.94 | 1.00 | 0.41 | 5.03 | 1.25 | 4.91 | 1.23 | 0.40 | 0.68 |
| 15. Data visualization design | 5.56 | 1.23 | 5.58 | 1.11 | 5.76 | 0.92 | 0.28 | 0.75 | 5.76 | 1.35 | 5.37 | 1.11 | 6.29 | 0.75 | 5.31 | 0.87 | 5.82 | 0.98 | 5.50 | 1.51 | 1.42 | 0.22 | 5.61 | 1.15 | 5.59 | 1.05 | 0.07 | 0.94 |
| 16. Monitoring design | 5.78 | 0.89 | 5.65 | 0.97 | 5.48 | 1.19 | 0.80 | 0.45 | 5.79 | 0.97 | 5.70 | 0.99 | 5.71 | 1.11 | 5.38 | 0.88 | 5.61 | 1.03 | 6.00 | 1.09 | 0.53 | 0.75 | 5.74 | 0.99 | 5.33 | 0.91 | 1.75 | 0.08 |
| 17. Service composition | 5.25 | 1.26 | 5.19 | 0.95 | 5.16 | 1.40 | 0.07 | 0.93 | 5.52 | 1.27 | 5.10 | 1.17 | 5.14 | 0.90 | 4.69 | 1.30 | 5.43 | 1.13 | 5.00 | 0.89 | 1.31 | 0.26 | 5.24 | 1.25 | 5.09 | 0.86 | 0.52 | 0.60 |
| 18. Event processing design | 5.54 | 1.05 | 5.74 | 0.98 | 5.32 | 1.14 | 1.25 | 0.28 | 5.62 | 1.26 | 5.68 | 0.88 | 5.57 | 1.27 | 5.06 | 0.99 | 5.54 | 1.07 | 6.00 | 0.63 | 1.04 | 0.39 | 5.65 | 1.08 | 5.14 | 0.77 | 2.1 | 0.03 |
| 19. Platform architecture design | 5.49 | 1.19 | 5.44 | 1.20 | 5.68 | 1.24 | 0.32 | 0.72 | 5.41 | 1.26 | 5.41 | 1.14 | 5.43 | 1.27 | 5.06 | 1.48 | 5.89 | 1.03 | 6.17 | 0.75 | 1.49 | 0.19 | 5.54 | 1.24 | 5.36 | 1.00 | 0.63 | 0.52 |
| 20. Smart object arch. design | 5.88 | 1.03 | 5.93 | 1.00 | 5.72 | 1.06 | 0.34 | 0.71 | 6.00 | 1.25 | 5.78 | 0.90 | 5.71 | 1.11 | 5.69 | 1.01 | 5.93 | 1.01 | 6.17 | 0.75 | 0.39 | 0.85 | 5.93 | 1.06 | 5.55 | 0.73 | 1.62 | 0.10 |
| 21. Interaction design | 5.29 | 1.23 | 5.26 | 1.02 | 5.20 | 1.41 | 0.04 | 0.95 | 5.45 | 1.12 | 5.12 | 1.18 | 5.29 | 0.95 | 4.63 | 1.31 | 5.00 | 1.23 | 5.00 | 0.63 | 1.96 | 0.08 | 5.23 | 1.21 | 5.41 | 1.14 | -0.44 | 0.52 |
| 22. Application arch. design | 5.64 | 1.22 | 6.02 | 0.85 | 5.36 | 1.38 | 2.84 | 0.06 | 5.83 | 1.25 | 5.56 | 1.05 | 5.71 | 0.75 | 5.75 | 1.34 | 5.75 | 1.29 | 6.00 | 1.09 | 0.26 | 0.93 | 5.70 | 1.21 | 5.82 | 0.95 | -0.44 | 0.65 |
| 23. Application coding | 5.71 | 1.09 | 5.74 | 1.04 | 5.88 | 1.30 | 0.2 | 0.81 | 5.90 | 1.37 | 5.44 | 0.86 | 6.14 | 0.69 | 5.38 | 1.31 | 6.18 | 1.05 | 5.83 | 0.75 | 2.19 | 0.05 | 5.78 | 1.10 | 5.64 | 1.21 | 0.55 | 0.58 |
| 24. Smart object coding | 5.78 | 1.11 | 5.77 | 1.06 | 5.88 | 1.23 | 0.09 | 0.91 | 6.17 | 1.19 | 5.61 | 0.97 | 5.86 | 1.06 | 5.69 | 1.35 | 5.75 | 1.17 | 5.67 | 0.51 | 0.94 | 0.45 | 5.81 | 1.16 | 5.73 | 0.88 | 0.31 | 0.75 |
| 25. Platform coding | 5.71 | 1.09 | 5.56 | 0.95 | 5.88 | 1.23 | 0.71 | 0.49 | 6.10 | 1.23 | 5.61 | 0.91 | 5.86 | 0.90 | 5.56 | 1.15 | 5.32 | 1.12 | 6.17 | 0.40 | 1.92 | 0.09 | 5.75 | 1.11 | 5.41 | 0.85 | 1.36 | 0.17 |
| 26. Test (all types) | 6.25 | 0.75 | 6.19 | 0.79 | 6.24 | 0.72 | 0.10 | 0.90 | 6.34 | 0.72 | 6.20 | 0.84 | 6.14 | 0.90 | 6.25 | 0.68 | 6.11 | 0.73 | 6.50 | 0.54 | 0.46 | 0.80 | 6.24 | 0.77 | 6.18 | 0.66 | 0.31 | 0.75 |
| 27. Installation (all types) | 5.58 | 1.08 | 5.77 | 0.99 | 5.76 | 1.30 | 0.46 | 0.63 | 5.93 | 0.96 | 5.66 | 1.19 | 5.57 | 0.97 | 5.69 | 0.94 | 5.50 | 1.26 | 5.50 | 0.83 | 0.49 | 0.78 | 5.76 | 1.02 | 5.27 | 1.35 | 1.92 | 0.05 |

## 3.3 Challenging process tasks in practice

**Qualitative analysis**. To analyze the responses to RQ3, we used our proposed framework to organize responses based on each task to deduce a higher level of abstraction of comments. In explaining perceptions and experiences, the respondents highlighted challenges related to the tasks making IoT development process different compared to non-IoT system development as described in the following subsections 3.3.1 to 3.3.3. For the sake of space, we cannot report all comments, as they are publically available at [46], and do not repeat similar explanations. We delineate the phases and tasks synthesized with the relevant quotes from both the survey participants, if any, as qualitative confirmatory evidence to answer the research question RQ3. Hence, our findings are contextualized, explaining why tasks are considered important and they raise issues during the development. In the following, we use icons to indicate, ❶ if a quote is about a problem, ☞ recommendation, and ❶ general information. We use an individual identifier and number P#, for each participants' quote.

### 3.3.1 Challenges related to analysis phase

The phase is mainly aimed at understanding what the IoT system will provide and, includes four tasks *ideation, requirements analysis, stakeholder analysis,* and *plan definition*.

### 3.3.1.1 Ideation

This task is to develop, refine, and prioritize an IoT system ideas based on problems from a business point of view and goals to be achieved via techniques such as brainstorming. Respondents found the *ideation* task a difficult exercise to stimulate free thinking due to unrealistic expectations and uncertainty of values provided by an IoT system:

> ❶ *Identification of what is the benefit of an IoT system is not easy... customers and developers might have unreasonable ideas... what can be done? what cannot be done? what can be done easier? [P9].*

> ❶ *Any IoT solution needs to clarify the features on the value chain of the need including application hardware integration sensor, electronics/ firmware integration, network management, software platform, data analysis, and added value for the end user [P59].*

> ❶ *Understanding business value for the relevant customers is a must but difficult [P86].*

### 3.3.1.2 Requirement analysis

Task *requirements analysis* has four sub tasks *domain requirements analysis, software application requirement analysis, smart objects requirements analysis,* and *infrastructure requirements analysis*. Forty respondents (40) stressed the elicitation and validation of the requirements as an important task. This rationale was captured quite well by a respondent:

> ❶ *Requirement is extremely important for development of IoT Systems, in terms of type of data required, frequency of data, use cases specific for business domain for which the platform needs to be built for and most importantly the end customer needs who will be the user of the system [P96]*

The respondents mentioned several reasons for difficulty in addressing various stakeholder's requirements as they dynamically change and new expectations arises as the system keep evolving:

> ❶ *Gathering the requirements and refining the requirements as you progress during the development stage never will be static. They should continue to evolve as assumptions you have made are proven false and/or in need of adjustment and new assumptions arise to replace and/or supplement the original set of assumptions [P83].*

> ❶ *IoT is in very early stage and usually use cases are not mature enough then requirements are not clear which might lead to multiple requirement changing throughout the project [P43].*

Different groups of stakeholders in an IoT project have different requirements to be identified and addressed:

> ❶ *As nature of IoT technology, every customer/industry has their own requirements, you should be flexible enough to meet every customers' requirements. To manage IoT projects, requirements of broader horizontal expertise on new technologies, e.g., electronics, communication, software development, etc should be identified [P7].*

> ❶ *Keeping requirements consistency is an issue [P3]*

In an IoT project, *groups of stakeholders that are from industry sectors have different requirements [P7]* where *keeping requirements consistency is an issue [P3]*.
An area of concern for the respondents was the difficulty of specifying security and privacy requirements related to storing personal or interaction data:

> ❶ *Most often system performance and capability could suffer due to satisfying the end user's privacy specific concerns. In our project it was hard to negotiate with common people and agree on what to capture and what not in private context of a home and resident where our smart home services may involve image capturing and processing [P2].*

When it comes to various requirements, the respondents also expressed the frustration about addressing contrasting requirements, such as scalability vs. data privacy:

> ❶ *Addressing the scalability requirements is a key to enable to compose multiple services as the number of users grows.... But we need to meet the privacy requirements for customer [financial] and sensitive data and scalability in our projects. This forces us to deployed system on servers located in our country region [P20].*

### 3.3.1.3 Stakeholder analysis

Identifying who are the right stakeholders to elicit system requirements form is critical to the success of any system development endeavor, and is equally important in IoT projects. IoT projects may take stakeholders from different industrial fields with different backgrounds as recognized by the respondents:

> ❶ *The industrial IoT space presents its own challenges as culture and backgrounds of people involved can be very*

*different [P44].*

❶ *The variety of stakeholders and customers that must work together [P47].*

### 3.3.1.4 Plan definition

Beyond conventional software system planning, such as project time frames and necessary resources, software engineers should take into account how the *interoperability* of heterogeneous components influences the project cost and effort required in further phases. A so-called *interoperability-based IoT project planning*, as also recommended by SCRM [S19] and BSI [S45], enables identifying interoperability requirements to be addressed in advance. This also was pointed by respondents, for example:

❶ *The project plan definition should be mainly dedicated to the interoperability as we need to integrate many devices of existing brands and our custom devices all under one IoT platform [P2].*

Some participants believed that the distributed nature of software and hardware components, as they have different independent development lifecycles, force to have supplementary plans for *testing [P5], [P11], [P15], [P16], [P33]*, and *deployment [P15], [P37], [P50]*.

What experiences exists amongst software engineers may not sufficient when moving to IoT projects. A mixed knowledge of software, hardware, and networking programming skills to design and implement at the different layers of IoT system deems to be difficult to find in software teams and impact coding productivity as was cautioned by respondents. Many respondents highlighted this issue during planning. Illustrating this, example opinions are:

❶ *A diversity of software development technologies required for developing a complete end-to-end IoT system... device development requires embedded software development skills, and low-level programming languages such as C. gateway development requires embedded application development skills… IoT backend system development requires cloud and backend development skills, mastery of various analytics technologies and/or machine learning systems. It is extremely difficult to find developers who can master all the development technologies in a large-scale end-to-end IoT system [P37].*

❶ *Identifying the right team to work on before developing an IoT project as issues detected at later stage could prove to be a costly and time wasting exercise [P75].*

❶ *IoT systems usually require a broader depth of domain knowledge and expertise. There is currently far less people with experience in developing IoT systems [P44].*

### 3.3.2 Challenges related to design phase

This phase is to exploit and map the elicited requirements from stakeholders that are identified in the previous phase to a to-be solution architecture which addresses the requirements via a collection of software application components, IoT platforms, things, and infrastructure. In ad-dition to the general architecture design tasks *service composition design*, *event processing design*, *resource discovery design*, and *monitoring design*, there are three specific classes of architecture design, i.e., *application architecture design*, *platform architecture design*, and *smart object architecture design*, that together provide a holistic architecture which guarantees flawless operation of system components (Fig. 8). Each is elaborated in the following subsections 3.3.2.1 to 3.3.2.3.

### 3.3.2.1 Application architecture design

A wide range of applications such as mobile applications, legacy applications, business analytics reports, and virtualization are designed to enable end-uses to use core functions provided by an IoT system. Although IoT application design has many similarities to conventional system application development, an important issue that should be borne in mind is the role of software applications to handle interruptions in hardware components, e.g., smart objects:

❶ *Smart objects have limited hardware resources. Any fault or failure at hardware level must be taken care by fault tolerance implementation by software, e.g. try to restart the device, add backup device, and continuously capture device state and data. For example, in smart home setting, if any device stops responding or lags communication, backup device must be contacted by control software/services [P2].*

### 3.3.2.2 Platform architecture design

As presented in Fig 9, there are seven preliminary subtasks, under the *data management design* task, used to design mechanisms to collect, clean, store, process, and visualize both structured and non-structured data from internal and external environments such as applications and smart objects. Respondents agreed the importance of the *data management design* task in an IoT system, but also tried to be cautious about the data quality concern influencing the development process cost. A respondent shared an overall comment about this tasks and its relevant sub-classes:

❶ *Removing nulls, multi-level cleaning logic for de-duplicating, data labelling, and some ad-hoc manual text manipulation all are challenging to perform, e.g. cost and effort, but these help to do further data processing task, i.e., providing personal finance insights for example [P30].*

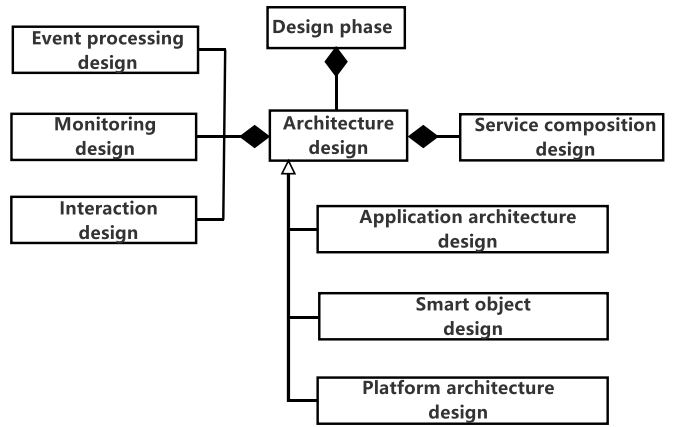

Fig.8. Tasks in design phase

❶ *The amount of time spent on collecting, cleaning, and labelling good quality data for model development is always underestimated in IoT projects. Our company was no exception, as considerable resources with our team were to prepare the data for model development for example identifying right product catalogues and labelling data [P42].*

An important issue when doing task *meta-data design*, i.e., defining mechanisms to facilitate the identification, classification, reasoning, retrieval, and exchanging of data between heterogeneous and distributed system components in adherence to interoperability and discoverability, is the fact that the operational environment of IoT system is changing. Related to this, the importance of designing dynamic meta-data for accurate data processing was another common issue stated:

❶ *The meta-data, such as ontologies, that we use for data processing should regularly be revised and updated to address the upcoming changing objectives of the IoT system and be aligned with application and stakeholder requirements. Data, i.e., catalogues of items in our projects that are scraped from various websites to use for training the deep learning model, are casually changing which our ontology is not able to detect them and to be used in our deep learning models. For instance, sometimes there is an inconsistency between new meta-data created by our users, i.e., merchants for item categorization, and our meta-data adopted in our platform [P30].*

Providing an effective data visualization including dashboards, reports, message boards, 3D spaces, and 2D maps, as defined in *data visualization design* task should be addressed in an IoT platform with respect to configurability and customizability, amongst other, in data visualization functions.

❶ *Visualization components as gauges, charts, maps, tables should be designed configurable and should allow you to change their data sources, visual representation, and to organize widgets into logical groups and layout. Moreover, dashboard templates allow you to reuse one configuration for multiple device dashboards [P30].*

### 3.3.2.3 Smart object architecture design

We frequently received comments from the respondents to incorporate factors such as *efficient energy consumption* [P32], *low memory usage [P7]*, portability and interoperability of devices [P2] into the architecture design for smart objects.

### 3.3.2.4 General architecture design tasks

The importance of other architecture related design tasks *service composition design, event processing design, resource discovery design, monitoring design,* and *interaction design,* as highlighted in the IoT literature, for example VITAL [S2], IoT-ARM [S8], FIWARE [S12], SPITFIRE [S20], ESPRESSO [S48], and iCore [50], are reminiscent of the architecture design in conventional software engineering. For example, the purpose of *monitoring design* task is to ensure both software and hardware IoT system components, i.e., applications, platforms, smart objects, and infrastructure, to keep track of system performance to identify anomalies, correlations, or similar patterns of divergence. Monitoring data can be system log such as operation history, energy log, resource consumption of smart objects deployed in different regions and their battery lifetime, network log, transmission queue size, the number of collisions, packet error rate and other critical networking statistics. The importance of *monitoring design* was described:

❶ *In dynamic IoT environment it is difficult to find if an affected device or software, we should always have live event in the microcontroller to check the status of all devices and do a proper action could be taken before any mishap [P2].*

The *interaction design* task is to identify and specify the message flow, interrelationships, sequential logic, and interaction protocols between the system components. The task illustrates how certain functions are governed by interaction protocols and accomplished in the context of the overall system. The complexity of *interaction design* task was frequently mentioned by respondents:

❶ *IoT development is all about interaction/collaboration of distributed components. There will be a lot of components depending on each other. It is a challenging task to organize the development in a way to deliver value fast and still keeping the overview [P82].*

The respondents elaborated the complexity of identifying interactions between system components by terms such as *interaction between low-resource hardware and software [P76]*, *interaction between device and the IoT-network [P76]*, and *interaction of 3rd party web services to distribute data [P81]*. Another insight that emerged provided additional evidence in relation to this task, i.e., identifying the potential communication barriers between system components in terms of security or the quality data exchange:

❶ *This task helps us to identify communication issues. Through the analysis of things, we discovered that our IoT system abilities are restricted by privacy settings of users' mobile applications. We also realized that there is an issue related to data exchange in the system, i.e., generating text from low-quality images can cause noisy input data that can affect data processing, model outputs and insights provided to users of the system [P30].*

❶ *Interaction design is a major part of development process to explicitly identify communication issues of portable IoT sensors and devices such as resource poverty, computation*

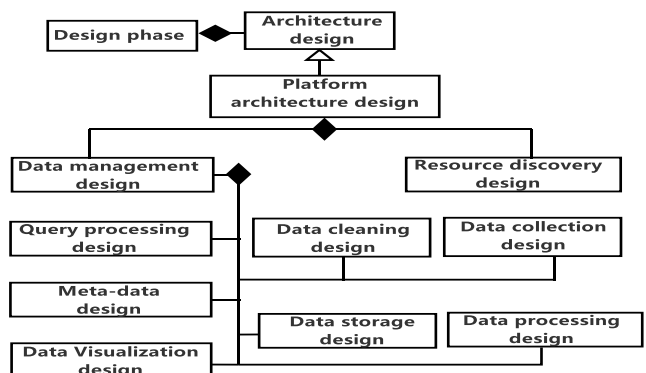

Fig.9. Sub-tasks related to platform architecture design task

*available on-device, device performance, and continuous. For example, traditional hardware systems, e.g. workstations and mobile devices, have bigger hardware sensors that ensure strong WiFi signals and reliability of data transfer. But, in IoT, due to weak capacity of GPS and WiFi sensors (miniature hardware), connectivity is a big issue as IoT device/sensor is prone to frequent disconnections and low transfer rates. All these should be identified and analyzed [P2].*

### 3.3.2.5 Other architecture design challenges

Some respondents commented overall challenges of *architecture design*. 30 respondents emphasized interoperability design for software and hardware components which differentiates the development of IoT systems from non-IoT ones. This is due to the lack of communication standards and protocols for the exchange and ingestion between software and hardware components which can dynamically join and leave the system. The interoperable architecture design for heterogeneous hardware components is the highest priority and an ongoing challenge:

❶ *There is no standard way to do integration. No standard models, protocols, and behavior descriptions. Different vendors try to keep their data and service in their own for business purposes. Many existing devices provide very limited service access through the vendors [P35].*

A similar expression, but from the software component viewpoint, was mentioned as well.

❶ *Integration of IoT system with enterprise system is critical. Looking at IoT system as connected devices and their data is wrong. IoT is an end-to-end system which leverage all existing systems, future systems and new services. For instance, IoT systems should be integrated with CRM, ERM, EAM, etc to triggers action which is critical [P51].*

Other than that, the respondents described the challenges of interoperable architecture design by terms such as *achieving interoperability in IoT systems is a difficult task [P7]*, *heterogeneity of data and devices [P29]*, *heterogeneous environment*, *integration of heterogeneous IoT devices and systems [P29]*, *device or things integration [P11]* and *communication protocol [P28]*.

According to the 19 respondents, the providing a design supporting a big connectivity and availability of different system components is one of the most challenging aspect of IoT *architecture design* task. Disconnection during execution time which may poses a huge risk for large-scale and critical IoT system happens for several reasons such network ban or broken network.

❶ *From my position, building IoT system, [for example], for the agriculture, most of the farms is in far ways with very weak connectivity and most of connecting ways are forbidden here. Wide-area network technologies like LoRa can be used in military purposes ONLY! [P49].*

❶ *Available network to guarantee the data collection, communication modules, and power consumption for IoT systems is the most challenging aspect [P64].*

### 3.3.3 Challenges related to implementation phase

This phase focuses on *coding*, *testing*, and *installation* tasks of IoT software and hardware system components, each with a specialized instance for applications, smart objects, and platforms, to operationalize the designed architecture in the previous phase. When we asked the respondents to specify technologies that they use during the implementation phase, as many as 127 respondents, 86 listed a variety of technologies (Table 4). The number in parenthesis represents the frequency of respondents who listed that technology. To be more specific, Arduino (%19), Microsoft Azure IoT (%17), Eclipse (%12), AWS IoT (%11.63), Raspberry (%9), IBM Watson (%5), ThingsBoard (%2) are the most popular pre-built IoT platforms using for both software and hardware components at the implementation phase. A minority of the respondents mentioned that they use Intel IoT, IoT Boards, Nokia Wing, and Vodafone INVENT (Table 3). Among the listed platforms, Eclipse and ThingsBoard are an open-source IoT platform for device management, data collection, processing, and visualization. Furthermore, the respondents reported the use of mainstream tools and programming languages such as Python, Java, C++, NodeJS, Matlab, Microsoft SQL, Microsoft Visual Studio, REST API, and React, for coding, scripting, transferring/streaming, and data manipulation. In particular, Node-RED and BoxPwr are programming tool for wiring together hardware devices. A few respondents (%2) relies on OMNeT and AnyLogic tools for compiling discrete event models and system dynamics simulation.

### 3.3.3.1 Coding

The issue of *integrity* of the development tool set for during *coding* tasks is not a trivial issue. Ranging from Python scripts for mobile applications, Scrapy library, to large platforms such as Microsoft Azure, to Azure Databricks, Azure Data Factory, and Scala, thus required integrity support. Some respondents mentioned frustration about raising incompatibilities with the version of same tools.

❶ *Sometimes libraries that we use are not compatible with latest version released and therefore lack certain features. For example, the Python language version on Azure has certain limitations for modelling activities [P30].*

This issue is equally a true fact for hardware components since the most effort in hardware coding is to provide a support for connecting a variety of third-party smart objects together. Illustrating this, one respondent said:

❶ *The integration of different third-party hardware each having different protocols and standards, make IoT implementation and testing difficult for us compared to other development types [P2].*

Despite the availability of a large suite of tools, some respondents believe that the *lack of right tools* is a challenging aspect. In relation to software components, for example, the respondents mentioned:

❶ *There are many IoT platforms and choosing the right platform (the trade-off of the ease of use and technical efficiency) in which you will collect everything together, and is the real challenge [P100].*

❶ *Use of appropriate tools and technologies to ensure QoS*

*for IoT [P12].*

❶ *Selection of most suitable code libraries/APIs [P19].*

This problem extended to hardware components as well.

❶ *The choice of sensors is critical for programming [P23].*

❶ *Finding the right mix of hardware components (underlying chipsets) is issue [P42].*

Clearly, the *choice* of software tools in implementation phase is undergone by the architectural design decisions in the design phase, which influences the tool integrity and subsequently software team productivity.

❶ *The coding activities are circumscribed by architectural decisions that are made earlier at the IoT design phase. This causes incompatibility between architectural choices and technologies needed for implementation. For instance, the choice of big data platform, i.e., in our case Microsoft Azure, restricted us to use of certain version of libraries when developing our deep learning model. Our development speed got down [P30].*

In contrast to literature suggesting that software and hardware components can be implemented via open source technologies [16], this was not the case where a respondent mentioned his tension to use well-established tools:

❶ *Open source tools specific to IoT (LoRA etc.) lack maturity. Well established tools and software services (Amazon AWS, Android Toolkit) can be effectively used to develop IoT systems [P2].*

On the other side of spectrum, the choice of hardware technologies influences the performance of software components. This rationale was:

❶ *A very well written IoT software applications may be influenced by the performance issues of IoT hardware whose state/data/ must be used and managed by software. This is why hardware with proper SDK/APIs/code libraries are almost preferred over ones that may be cheap, easy to implement and configure but lack sufficient software libraries [P2].*

### 3.3.3.2 Testing

Although testing tasks in IoT system development have many similarities to traditional system development, they become challenging as there is no a clear line between the distributed, heterogeneous, and strongly connected software and hardware components of IoT systems. Seventeen (17) respondents jointly shared their opinions about the testing issues in terms of its prolongation and the heterogeneity of components:

❶ *There might be many types of hardware and software components that should run harmoniously. Creating a testing environment might be difficult [P5].*

❶*You want something to last five years on battery but can't test it for five years to make sure it works. You need to do a lot of testing and put the devices through a variety of situations to ensure longevity [P32].*

❶ *The acceptance of the solution by the target audience, despite of all validations and due diligence, takes a lot of time [P46].*

One unique aspect of test in the IoT development process is the *interdependency* of test for hardware and software components, where each component is provided by external vendors, causing delay in timely and cost effective delivering system increments.

❶ *You need the hardware/firmware design ready to go to start implementing prototypes while your build a small percentage of the software. You can't complete software testing/regression testing until the hardware is in a testable and reliable state which takes month of iteration in itself [P48].*

Due to distributed nature of IoT-based systems where each component may be deployed in different geographical regions it is difficult to draw a line between the *test* and *instalment* tasks.

❶ *In contrast to theoretical SDLC, iterative, Agile methods, where testing and deployment could be considered as two distinct phases, in IoT context, actual testing mainly happens post-deployment. It will be more accurate to say that in IoT life engineering cycle, it's a single phase that primarily refers to test-driven deployment. For example, in logistics tracking, RFID tags that worked well as part of controlled experimentation exhibited totally different behaviours [due to differences in allowed frequency in different countries, herein United State and Pakistan. Software services and modules worked all fine in Pakistan however, due to regulatory issues of allowed frequency on which IoT devices (RFID sensors) could operate in United State, the software functionality needed to be rewritten to deal with low frequency issues [P2].*

TABLE 3 Technologies used by respondents during implementation phase

| Multiple occurrences | |
| --- | --- |
| Arduino (16) | Azure IoT (15) |
| Eclipse (10) | AWS IoT (10) |
| Raspberry (8) | Agile (8) |
| Python (5) | Matlab (5) |
| React (4) | IBM Watson (4) |
| NB-IoT (2) | DevOps (4) |
| ThingsBoard (3) | REST API (2) |
| Visual Studio (2) | ARM Embedded (2) |
| Buildroot (2) | |
| Yocto (2) | |

| Single occurrences | |
| --- | --- |
| Java | Master of Thins |
| Node-Red | Thing Workx |
| ARM Embedded | IoT Architecture Blueprint |
| Intel IOT | Azure IoT Hub |
| IoT Boards | Microsoft SQL |
| Nokia Wing | Particle |
| Isar | Cmake |
| MQTT | Django |
| HTTPS | AEPs |
| PTC ThingWorx | GDSP |
| IoT Hub | Event Hub |
| RTOS | Cervello |
| ESP8266/ESP32 | NodeJS |
| OMNeT | BoxPwr |
| AnyLogic | C++ |

The severity of incompatible regulations, like national communication protocols, across different countries during performing test, which appear to be costly, is also salient:

> ① *The software and hardware that worked well during testing was required to be altered post-deployment. In this case, regulatory issues concerning the IoT hardware, as per US laws, triggered the need to refactor some of the software coding to support new device frequency indicated by United Stated's law. We can say that hardware implications enforced changes in software modules for compatibility issues. In a general context, regulatory issues of IoT signalling and frequency also impact IoT source code and implementation [P2].*

Along this line, some respondents hasten to note the challenges of testing hardware components, for example smart objects that are deployed far from software teams and may not be always proximate and easily accessible to test and fix:

> ❶ *Conventional software systems do not involve physical objects which have to be deployed in the field for testing in order to get the software to actually work properly in the real world [P53]*

> ❶ *Debugging devices which stop communicating all suddenly, need for physically reaching the site for debugging device issues [P73]*

> ❶ *Quality testing IoT devices is crucial - once they are out then it is hard to fix [P61]*

### 3.3.3.2 Installation

Compared to the most non-IoT systems deployed on centralized host environments, IoT systems are deployed on distributed and decentralized network. The respondents raised the challenge of *installation* task due to complexity and increased the remotely deployed things:

> ❶ *One of the biggest challenges in IoT is deployment in the large, i.e., deploying and dynamically updating IoT systems that consist of thousands or even millions of things, e.g. sensors [P37].*

Other than that, some respondents believe that conventional software systems typically do not involve physical objects which have to be deployed in the environment. They highlighted challenges associated with physical devise installation by the terms such *water/rain proof electronic devices [P16]*, *installation of the device must be secured and fast [P23]*, *controlling noisy sensors and inaccurate motors [P46]*, and *ensuring the deployment doesn't become an Island of things by engagement with the right stakeholders [P103]*.

### 3.4 Recommendations of IoT developers to address outstanding software engineering challenges?

Based on their experience, the respondents shared some overall advices for software teams in addressing the discussed challenges in sections 3.3.1 to 3.3.3. In response to RQ4, we classified the recommendations under nine groups as synopsized in the following.

**Multi-disciplinary process.** The respondents pointed the fact that the development process of IoT systems, unlike other systems, is highly interdependent and tightly coupled with different facets such as market, cost, and technologies that account for consideration:

> ❶ *In most cases developing IoT systems force a multi discipline approach. There will always be at least driver code written for an embedded device as well as some web based development. Market pressure also means that Android and or IoS will need to be considered. This means your starting point has a minimum of 3 or 4 frameworks to contend with. Additionally, to make IoT devices affordable it's is likely that the team will need to undertake some electronics design and manufacture tasks [P1].*

> ❶ *IoT is not software development only. It needs a deep understanding of civil/ mechanical engineering, electronics, and radio frequency then properly integrated into IT solutions [P59].*

Changes in organizational structure, in development team and users, should also be taken into account:

> ❶ *The impact of IoT system requires organizational and mindset changes. Every one of the development team will be affected. From the developer, over the product manager, salesperson, till human resource [P82].*

IoT development has the learning curve with knowledge in both hardware and software components. Training and tutorials are needed to assist the learning of software engineers and to utilize community Q&A sites.

**Co-operative process.** IoT systems are typically developed jointly by multiple development teams and stakeholders. A good cooperation between these entities needed:

> ❶ *For IoT, the main reason for failure is because IoT projects are driven by IT team. This should be more collaborative and driven from the field experts such as civil, mechanic, instrumentation, etc [P59].*

> ❶ *An IoT project requires the cooperation of many contributors such as hardware manufacturer, network engineers, developers, clients. You must therefore work with all of them at the same time to make sure your IoT solution will be successful [P45].*

**Bespoke process.** The respondents expressed the necessity of having tailored development process aligned with project situational factors which can be categorized broadly as technical, (the choice of smart objects), organizational (project goals, or human), level of expertise, and privacy of end-users. There is no reason to think that the development process tailoring, which has a long-standing acknowledgment in the software engineering [14], is not required in IoT projects.

> ❶ *Development tasks depend on the situational factors. For instance, if a team's domain expertise is low, domain analysis becomes more important. It might be better to try to find out the situational factors that are central to*

*IoT system development. Afterwards, an importance/relevance ranking can be done according to the changing situational factors [P5].*

❶ *Development method or tool usage are dependent on the project or delivery at hand [P13].*

💡**Agile-driven process.** Incorporating *Agile software development practices*, *lean software development* and *iterative development*, i.e., think big and start to develop in gradual and iterate way, which are based upon Agile manifesto, into the development process of IoT system lessen the issue of uncertain and changing requirements:

❶ *IoT application requirements are more speculations when you write them down since the marketing and support team joins the project usually at a much later phase. Due to the high number of parties that must work together in realising an IoT solution, you should switch to a stage gate Agile methodology to enable you to get all key-decision-makers as early as possible on board and you benefit from their knowledge from a very early stage of the development [P45].*

💡**Secure-aware process.** Through the respondents' comments, there was a constant recommendation on addressing security requirements throughout the development process lifecycle and across the IoT system layers. Forty-five (45) respondents reckoned strict security aspects such as design security for everything, such as data streaming, sensors, device, communications, network, configuration, and user interface that should be identified at the early phases and governed during the development process lifecycle for all hardware and software components. The simplest way to explain the difference between security design in IoT systems and conventional system development was distinguished in terms of scale and variety of working things in the system:

❶ *When it comes to IoT, a very important but perhaps an overlooked aspect of designing IoT solutions is the selection process of the things regarding, among other criteria, security. Weak security practices on the side of the devices and their connectivity layer pose a huge risk for large-scale and critical IoT-based solutions [P63].*

❶ *Ensuring the data security and privacy particularly in countries where these are sensitive/enforced should be considered [P10].*

❶ *In IoT development we have to deal with huge amounts of data, intense communications between devices, vast range of user interfaces and security issue is different from conventional applications [P43].*

💡**Test-driven process.** Software engineers need to be prepared to conduct rigorous and concurrent testing of hardware and software components to ensure the code quality concerns such as the limited memory of device, massive data, and noisy environment, security, interoperability and performance. Otherwise, neglecting code quality is hard to rectify:

❶ *Developers are used to building software. In their mind, it's just code. However, when you throw real*

*physical hardware that is not commonly used and may have unexpected behavior introduces challenges that a developer has to consider early on. Some hardware issues can potentially brick a hardware if the software design is flawed and at this point software developer may not be able to bring the device back [P48].*

Not only common code refactoring techniques should be used to improve the system source codes, the behavior of the system upon external input events should be analyzed to further improvement of system source codes.

❶ *Dynamic analysis of source code refactoring needs high quality attention [P14].*

💡**Integrated architecture-driven process.** IoT development needs multi-level development processes such as embedded software and hardware components, each is governed by different teams who focus on only one group of architecture design tasks, but, these processes are not in isolation and have impact on each other. An integrated view- *co-design-* of software and hardware architecture design, due to reasons such as low-speed hardware and fast-speed software, is recommended.

❶ *There is an additional hardware module. A multi-code developer is needed and who knows how to handle both hardware, and software environments…it is necessary to have an integration of both environments [E88].*

❶ *IoT should always be seen from two perspectives, i.e., hardware and software. And, hardware should be seen through its phases - hardware design, writing firmware code [P42].*

💡**Automated processes.** One unique aspect of IoT systems is the right choice of compliant tool sets to lessen the effort for multi-device coding and testing hardware and software components and to achieve automation resulting development productivity:

❶ *Development of IoT based systems need to consider the degree of the automation [P24].*

💡**Energy-aware process.** Eleven respondents (11) recommended visibility into energy usage that emphasizes the graceful power optimization and factors such as event frequency and available resources at the all software and hardware levels of system development hierarchy. This is not limited to the design and implementation phases but also the analysis phase where the limitations of hardware components and the trade-off among cost, size, and reliability:

❶ *A lot comes from restrictions of hardware. Low memory, low power, and long life all provide their own unique challenges. For example, you are developing a software chat service which allows users to message other users with minimal delay and near-real time notifications when a message is received. In IoT land, this is very different. If you want to talk to a sensor, it may not always be able to listen for new messages and responding instantly. This is due to a decision made to reduce power consumption [P32].*

❶  *Managing the power for remotely deployed things [P7].*

Table 4, synopsizes the research findings –publically available at [46]– in the form of Five Ws Scheme (5W2H model), including when, what, why, how, and who (who refers to development teams) action plan. This would pave the way for designing engineering methodologies in the view of Whats (challenges) are associated with the tasks and How (recommendations) are available.

TABLE 4 Summary of Findings

| When (phase) | What (challenge) | Why | How (recommendations) |
|---|---|---|---|
| Analysis | • *Ideation* task: Uncertain value of IoT system | • Understanding attainable business values for an IoT system might not be perceivable by all stakeholders (e.g. *"an IoT toaster is nice, but does it solve any issue?"* [P12]) | • *Value creation*. Identify, analyze, and get agreement on alternative value chains achievable by an IoT system. |
| | • *Requirement analysis* task: Dynamic, conflicting, and evolving requirements | • There are far more unknowns in IoT systems as they involve a wide range of stakeholders, such as hardware/software development teams, end-users, and citizens with uncertain requirements and priorities. | • *Agile-driven process*. Incorporate agile practices such as lean governance, risk-based milestones, short iterations, and value people over process/tool into the IoT software development process. <br> • *Bespoke process*. Define bespoke development processes with respect to the choice of hardware/software technologies such as devices, platforms, programming tools, APIs, project goals, and environmental/organizational settings. The impact of tailoring should be assessed and discussed by development teams. |
| | • *Plan definition* task: Variety of required skills | • An IoT system relies on diverse technologies such as embedded software development, low-level programming languages such as C, backend programming, device configuration, data analytics, and machine learning algorithms, and so on. It may be difficult to find teams with sufficient mastery over these technologies. | • *Multi-disciplinary process*. Create a cross-functional team and develop collaborative communication between hardware and software teams and experts from different fields of experts in the project domain. For example, if the IoT system is developed for the healthcare domain, bring experts and ask for their feedback. |
| | • *Requirements analysis* task: Data security and privacy | • Different stakeholders and a wide range of heterogeneous hardware/software components are joining and leaving an ever-changing ecosystem for IoT systems. Each brings with it different security concerns. | • *Secure-aware process*. Define security and privacy requirements and mechanisms to monitor and control software and physical objects throughout the development lifecycle and system layers. |
| | • *Stakeholder analysis* task: Stakeholder collaboration with different industrial and cultural backgrounds | • Stakeholders from different industry domains, cultural backgrounds, and skillsets are involved in an IoT software development project. | • *Stakeholders-driven process*. Identify and engage the right stakeholders that are important and specify their responsibilities during the development process of an IoT system. <br> • *Training*. Familiarize both hardware and software development teams and stakeholders with the different aspects of IoT system development and interdependencies between hardware and software components (e.g., 1-day in-house tutorial, Web research). |
| Design | • *Architecture design* task: No standardization for data ingestion and exchange between hardware/software components, heterogeneity of data and devices, and communication protocols | • IoT third-party hardware/software components use their own protocols, data formats, communication models, and semantics to exchange data and call their functions. | • *Integrated-architecture driven process*. Identify interaction points between software and hardware components and define mechanisms such as adaptors or proxies for interoperable plug-and-play components to communicate. Note that hardware interfaces do not change as fast compared to many software components. <br> • *Interoperability plan*. Define a plan with a focus on interoperability including integration strategies and their costs which used in the design phase. |
| | • *Architecture design* task: Unpredictable disconnection and unavailability of distributed software/hardware components, e.g. third-party | • Components may become unavailable due to device battery/power run-out, Internet/wireless network restriction access or regulations | • *Replicate-driven design*. Use common system engineering practices such as replicating and synchronization of hardware/software components of IoT system based on the different layers of |

| | smart objects devices, and legacy systems | | IoT chain. |
|---|---|---|---|
| | • *Architecture design* task: Resource poverty of IoT sensors/devices | •Limitations of hardware components and technologies chosen or enforced by third-party developers' decisions to reduce power consumption. <br> •A trade-off between portability and context-sensitivity that inherits resource poverty | •*Power-aware architecture design.* Adopt graceful power optimization techniques for (remotely) deployed things in architecture design. |
| Implementation | • *Coding* task: Incompatible programming libraries, third-party devices, communication protocols <br> • *Coding* task: Lack of maturity or right tools and choice of smart objects, e.g. sensors <br> • *Testing* task: Debugging and configuring | • Third-party coding libraries and hardware technologies have their own limited APIs and predefined protocols which might be incomplete or inconsistent. <br> •Remote and low-memory hardware components in noisy environments | •*Tool selection.* Consider the choice platform, hardware technology, tool, APIs, visualisation using KPI, reports, graphs, and dashboard. <br> •*Dynamic code analysis and refactoring* <br> •*Automated processes.* Increase automated coding and testing system components. |

# 4 DISCUSSION

## 4.1 Implications for research and practice

*Research implications*. There are implications that arise from this research. Although debatable, we believe that IoT development process is not necessarily a new revolutionary genre and it can be well connected and positioned in the conventional wisdom of software engineering, however, it reappears taking to an extreme level. We observed such an interpretation in some respondents' comments in relation to RQ2 who speculated that IoT development processes possess many issues, traits, and principles of conventional system development process, but, issues are scaled up in IoT context. That is why 3 out of 127 respondents did not believe much noticeable distinctions in IoT development process challenges compared to the conventional software development, for example:

> ❶ *In my perspective, there is not much of a difference. IoT is just a new term used for hardware components that can connect, interact and consolidate information to be perceived and monitored. Things existed in IoT sense even before. e.g., wireless devices talking to each other over larger regions using long range relays [P42].*

Secondly, it is important to realize inherent challenges at the micro process level, i.e., performing development tasks, signifying that not everything necessarily from the conventional software engineering literature can be equally applied in an IoT context. In IoT projects, software teams deal with very large amounts of data, design interaction between software and hardware components, cooperation of different stakeholders, vast range of user interfaces, and security and reliability issues. For example, as discussed in Section 3.3.1.2, identifying rapidly-changing requirements from various key stakeholders in IoT is a challenging exercise compared to conventional software engineering. This is also true, as noted earlier (Section 3.3.4), for a security-aware development process which extends concerns from software components to hardware ones.

The third research implication is the common intersection between the findings in this research and those that have already highlighted in other software engineering domains. While our findings (Section 3) have originality for software engineering researchers, they are in conjunction of the results in other software engineering domains, which increases the credibility of our findings. For example, the framework tasks under *data management design* task lead to the quality of IoT-based systems, which concur with the findings of Kim et al. [19] reporting on challenges of software engineering of data analytics systems. Our findings (Section 3.3.2.3) are confirmatory to their work as both studies mention the importance of data quality design. In line with this, as IoT systems have share issues such as heterogeneously and security of components, as presented discussed earlier, with other modern computing paradigms namely cloud computing and data analytics. Consequently, the findings in Section 3 can be exploited to inform software engineering researchers about the development process of cloud-based or data analytics applications.

*Implications for practice*. Our proposed framework is important for software practitioners in a few ways. Firstly, it can be as a guideline for software engineering managers showing tasks that are anticipated for IoT scenarios. Based on the personal experiences of our respondents, the framework has been enriched with a list of challenges and recommendations that software engineers should be aware of. For example, in Section 3.4, we provided some recommendations such as the choice of development tools, frequent and dynamic code review of software and hardware components, cultural shift, security, and training for the inclusion in the development process. These would pave the way for the design of new IoT specific independent and decomposable IoT software engineering methodologies. Secondly, software teams can treat the framework as a tool or to-do checklist to appraise the extent to which their in-house development approach supports the development of IoT systems. They can use the framework's tasks in their development approach to extend its capability for IoT development. Moreover, as the framework includes the domain-independent tasks, it can be used as an evaluation tool to identify shortcomings, strengths, similarities, and differences among alternative approaches for IoT development. In other words, the framework aids the selection of approaches that suit situation-specific characteristics relevant to a given IoT project. Such a normative application of the presented framework, is in line with the *software process improvement* as quality management initiatives committed by software teams to ensure quality and repeatable development processes in a cost-effective way, herein IoT systems. Thirdly, despite our reported recommendations in (Section 3.4) being IoT centered, there is no reason to think that they cannot be applied to system development processes for other computing paradigms. Software teams can examine if the findings of this study, the development process tasks and identified challenges, will reoccur.

## 4.2 Research limitations and future work

The results of our research need to be understood in light of some limitations. The framework definition has been abstracted away from underlying technical and domain-specific implementation details. We did not discuss how the framework is formally integrated into existing in-house development processes. This, foremost, needs to define technology-centric techniques and heuristics to operationalize our proposed conceptual framework per task. Future research is needed to properly enrich the framework via implementation techniques.

When software teams integrate development processes, they may consider who is responsible for the execution of each framework task and associated steps. In addition to conventional software engineering roles, IoT specific ones, if any should be defined, which is, yet another, the important subject of future research. For example, a stakeholder-driven IoT system development.

Our paper solidifies a conceptual framework, including phases and tasks, supported by domain experts' opinions. The presented recommendations in Section 3.4 do not supplant the engineers' expertise and subjective judgment about which task set should be performed for an IoT project. As pointed out by the respondents in Section 3.3.4, there is a need for tailoring of development process with respect to IoT project situational factors. Otherwise, there is a risk of losing project goals. This raises questions about how developers select or tailor development processes for IoT context. This is in line with the findings in Fahmideh et al. [5] suggesting to design situation-specific IoT development approaches, which is reiterated by our survey respondents. Development teams can use a wide range of approaches like assembly-based method engineering, [47] ,[48] and the semi-automated way [49] discussing mechanisms to deal with this through the construction of project-specific development process which can be equally applied in IoT domain. The presented framework in this paper is a starting point for studying how development processes can be tailored for IoT system development projects.

## 5 THREATS TO VALIDITY

There are threats to the validity of this research's findings, based on the conducted research method (Fig 1.), in terms of *construct validity*, *internal validity*, *external validity*, and *reliability* [50] which is discussed in this section.

### 5.1 Construct validity

*Construct validity* concerns the adequacy of measures used to test a concept studied [50], herein the framework's tasks. We defined a process framework in an exploratory fashion that is rarely explored in software engineering. The question is therefore whether the tasks are representative of a development process lifecycle for IoT systems. We found that the phrase *IoT development process*, with its multi-disciplinary nature, is a new topic and interpreted differently by people in different domains. During phase 1, we did not find any clear or universal definition of the term and hence the presented framework in Fig. 7 is a cumulative viewpoint of different studies. This may raise the concern as to whether we have correctly measured the validity of the framework and if the right questions have been asked from participants in the right way. To minimize this, in Section 2.1.2, we discussed the issues of defining the appropriate granularity level, an aspect of any conceptual modelling endeavour [17],[37], and the fact that we tried to identify common recurring tasks that are stated in the literature for this framework. Since survey participants gave promising comments about the survey questions, i.e., representing the framework, this threat can be considered mitigated.

Our adopted mixed methods research in this paper is informed by prior works [17],[37] recommending two high-level exploratory and confirmatory phases to get an in-depth understanding of a phenomenon that is in its infancy stage. Second threat to construct validity with this type of research method is related to the dependability of the research phases, i.e. framework derivation and framework evaluation. As both phases have been conducted by the same research team, there is a concern that the researchers' opinions may have biased the research findings. To minimize this effect, we engaged in a technique termed *venting* [51] or *cross-checking* [52], through which we informally discussed the framework and piloted our draft survey instrument in multiple occasions with practitioners and researchers who were interested in IoT to get us their opinions that allowed us to refine the framework and survey. For example, one reviewer suggested us to visualize the framework in the survey and some clarifications to phrase the tasks to present clear items to be evaluated in the survey by the respondents. During the survey data collection, we included questions that asked respondents to share their observed critical issues about the presented framework and our survey instrument. In one particular instance, out of 127 respondents, two commented that providing situational factors indicating whether incorporating a framework task is a missing feature of the framework. In line with this comment, three respondents added that rating the importance of tasks in the survey might not be a plausible question due to requirements variability in IoT project. One respondent believed that our research question cannot be examined because *"IoT projects are not typically approached with such formal framework... everyone is making it up as they go… compare IoT to web development in the '90s. It's all new and unknown."* Except for this opinion, we did not receive any fundamental critiques that could introduce a serious construct validity issue with our research method. Furthermore, from our surveyed industry experts ranging from managers to software/hardware developers located in different countries and in different IoT development domains, we found that the framework and survey instrument are resonated with respondents' experience and no fundamental critique related to the research phases was reported. Finally, the post-hoc analysis with the alpha = 0.05, effect size d = 0.2, and statistical power 0.7, as reported in Section 3.2, indicated that our sample size 127 is sufficient to identify a given effect size at certain percentage, i.e. %99.

### 5.2 Internal validity

As for *internal validity*, i.e., concerning with situations that may have affected research outcomes whilst researcher has not been aware of them [50], we cannot ignore the fact that our time/effort intensive survey needed the participants' concentration to respond the questions accurately. The issue of respondent fatigue bias may have caused some respondents to complete the survey in a cursory way, which may have had a negative impact on the rating of the framework's tasks or open-ended questions and thus limited the validity of the reported statistical tests and quotes. In all cases that we found odd or having unclear responses in an answer sheet, we directly contacted the respondents to double-check the accuracy of her/his responses. A further issue related to the internal validity is the survey design, in particular, misunderstanding the survey questions due to the complex and multi-faceted nature of IoT systems. As a countermeasure, during our purpose sampling, we explained—both verbally and in writing—the terminologies and objective of the survey for the selected participants to give them an understanding of our research objective and our perspective with respect to IoT system development. Additionally, although it was mentioned that IoT system development comprises both software and hardware teams, the number of survey participants from software groups was significantly less than hardware teams (93 vs. 3), as discussed in Section 2.2, this may have caused a biased view of the investigated framework. A good frequency of comments related to hardware aspects of IoT development, presented in Section 3.3, likely lessened this issue.

### 5.3 External validity

With respect to the *external validity*, i.e., the extent to which research outcomes can be generalized to other contexts, firstly, in the derivation of the framework, we tended to identify commonly grounded tasks in the literature. Nevertheless, it is likely that we have missed other, yet less cited, tasks that could be important for the inclusion in the framework. We do not claim that the framework is capable to manifest all necessary tasks for all IoT system development scenarios. Moreover, as we purposefully selected random samples for the survey, the survey findings may be limited to those who participated in this survey and generating biased results. As discussed in Section 2.2, whilst we experienced the lack of expert participants, due to infancy stage of IoT field, we tried to have participants from a range of backgrounds in our limited sample size, which is in line with the suggested maximum variation sampling strategy by Patton [41]. We abstain from claiming the generality of our research findings. However, in light of maximum variation sampling rule [41] that suggests varying opinions and experiences to avoid attrition bias, we ensure that the respondents from 35 countries distributed across 6 continents, 15 industry sectors, seniority levels, and years of experience are good representative to answer to the stated research questions. Finally, we needed to identify experts who would be active in IoT system development to participate in our survey. We contacted many experts and development companies but only hearing that they could not share their experience with entities external to their companies due to privacy, regulatory issues, and intellectual property matter. We are aware that some challenges related to the tasks may not have been shared by the respondents or have been misrepresented. Thus, we cannot firmly state that the framework and findings from our survey in this article are completely the representative of the opinions about IoT system development process.

## 6 RELATED WORKS

The IoT field is polymorphic and multidimensional but in this research we narrowed our focus on the development process for IoT systems. Given that, we divided the literature of software engineering for IoT into *less-related* and *closely-related* studies are delineated in the following subsections.

### 6.1 Enabling technologies for IoT systems

The less related literature refers to enabling technologies for IoT systems. These are the synergical Internet-based computing paradigms such as cloud computing, data analytics, and blockchain that have similar characteristics and provide backbone services for IoT systems. For example, cloud computing primarily helps IoT via providing resources for the storage and distributed processing of the acquired sensor data. We reviewed the existing process frameworks related to cloud-based system development to find a possible intersection with our proposed framework. Via adopting a metamodeling approach, Fahmideh et al. [17], propose a generic cloud-centric software application development process framework, including 18 tasks identified from 75 papers, which integrates various viewpoints of the same development process of cloud-based systems. We found that blockchain technology is attributed to empower IoT systems for the security of exchanging data between hardware and software components. The survey of 156 blockchain developers by Bosu et al. [18] is to identify the differences between the development processes of blockchain-based systems and conventional systems. Our work focuses on the development process tasks and their related challenges while Bous et al. focus on the motivations of developers towards this class of system and challenges related to tools. All of the abovementioned works are silent to uncover the IoT-centric development process as discussed in this paper.

IoT systems rely on computing capabilities of data analytics systems. In a survey, Kim et al. asked data scientists of challenges they face and practices use to handle these challenges [19]. Based on respondents' professional experience, skills, education, working style and time spent on work activities, the authors clustered the survey respondents into three categories named data, analysis, and people. While both our work and Harris et al. use a survey-based research approach, our focus is on the development process. Regarding the challenges that both IoT and data analytics systems may face, both studies point the issue of data quality, implying how prior work in other domains can, in conjunction with our findings, stimulates further studies.

### 6.2 Software engineering for IoT systems

Few studies to date deal with the software engineering methodology attuned to IoT system development. An issue with methodologies like Collin's [20], Ignite [21], Inter-Meth [22], ACOSO-Meth [23] is that they are too general and largely untested in practice. Another limitation of these works is where they express the same development process but in different ways and levels of abstraction. Whilst it is unavoidable fact to have a variety of methodology designs for an application domain where each has its own focus and scope, an overall domain-independent and empirically evaluated framework by domain experts that highlights key development process tasks, as reported here, does not yet exist.

Apart from the multitude reported ad-hoc techniques for addressing challenges of IoT systems, but applicable at the implementation level, such as programming of heterogeneous objects [53], seamless integration [54], or platform development [55] to name few, there is a stream of research that focuses on accommodating of common software engineering approaches. These includes mashup-based (WoT-Kit [56]), model-driven development (ThingML[57]), domain-specific languages (Midgar [58]), or visual programming languages ([59]) that facilitate the implementation of both software and hardware components of IoT systems. These studies concern topics such as service composition, automated code generation, and addressing the interoperability of things. In this spirit, Ciccozzi et al. [60] propose a model-driven engineering methodology with a focus on addressing the issues of interoperability among components of real-time IoT systems. Nevertheless, one finds it difficult how their suggested fine-granular techniques can be pulled up together to form a series of high-level intellectual bins, i.e., the conceptual process framework to provide a full end to end picture of developing IoT systems. We believe that such implementation techniques can be greatly adopted to operationalize the framework. In addition to their platforms in marketplace, big players like IBM [S47], Cisco [S6], Amazon [S64], Google Cloud IoT [S65], CityPulse [S39], ThingSpeak [S59], and Microsoft Azure IoT [S66] (see Appendix A) provide some case studies of successful implementation of IoT-based systems, however, there has been little attention to underlying development process enacted. Whilst the presented process framework has been inspired by existing work in the literature, we extended them with new important features in the following ways:

*Focus and depth of analysis*. In contrast to previous research, our study is the first one that explores the aspect of the development process for IoT software systems. It narrows its view to existing proposals providing either a complete or a partial approach for the development of software for IoT platforms. Thus, it is much more focused compared to the abovementioned works. Our proposed framework encompasses key tasks, challenges, and recommendations which can be sequenced into the existing fragmented structured system development processes to enhance their capability to support IoT systems. Moreover, the framework can be used as a checklist to assess the level of support offered by existing approaches for IoT system development. Furthermore, providing a deeper explanation of the framework organized into tasks under different phases enriched with IoT experts' opinions is another distinct feature of the current study. Such features have not been covered by any of the existing works.

*Research method*. In comparison to prior studies, our proposed framework has been based on the use of explorative research techniques including qualitative and quantitative ones which have been unavailable in reviewed works. As noted in *"Section 2.1.2 Steps for framework derivation"*, the framework derivation has been based on (i) top-down way where reviewing general literature on IoT and conventional software engineering gave us insights on the foundation of IoT system development (such as basic components devices, smart objects, systems, and platforms as noted in Section 2.1.1); and (ii) bottom-up way via examining and abstracting out different existing studies focusing either partial or fully on IoT development process. Moreover, our study has been done at a large scale of evaluation by a great many IoT experts, as opposed to existing studies were their findings are quite limited from an empirical evaluation perspective.

# 7 SUMMARY

The objective of this revelatory study was to demystify the software engineering of IoT software systems from the perspective of the development process. This is a substantial factor for development teams to achieve more maintainable IoT systems. We presented the first in-depth study to answer this issue by identifying a process framework including 27 development tasks organized in a three-phase based process framework and then obtaining quantitative and qualitative support through a Web-based survey results. We provided some recommendations to manage the development process complexity of this class of systems. The software engineering literature has not observed such a scientific and contextual understanding of the overall development process as it is anticipated to get to know what is new or re-iterated.

This research and its explanatory generic framework supersede or will ultimately become commonly accepted by researchers is a lofty goal. What is significant is that the framework was derived through synthesizing the literature pertaining to the IoT field and validated it in a stepped and transparent way. Our findings have implications for both research and practice:
• It envisages a codified development process for IoT systems in the words of software teams along with given concrete examples grounded in their experience. Given the current lack of understanding on this topic, the framework provides an overarching view helping to better focus and classify future research studies, and, hopefully, stimulates researchers to ask further research.
• It represents a starting point for improving research and practice understanding how IoT systems can be better planned, anticipated and organized from the perspective of software development process.

## ACKNOWLEDGMENT

The authors thank the generous feedback and helpful suggestions, especially from associate editor and reviewers. The authors would like to thank Dr. Christopher Magee, the associate professor from the Faculty of Social Sciences, School of Psychology at the University of Wollongong for his constructive comments and suggestions for improving our data analysis. Finally, Professor John Grundy is supported by ARC Laureate Fellowship FL190100035.

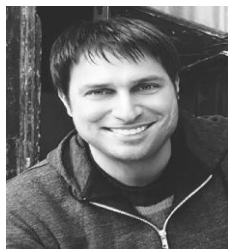

**Mahdi Fahmideh** is an Assistant Professor at the School of Computing and Information Technology, University of Wollongong, Australia. He received a PhD degree in Information Systems from the Business School of University of New South Wales (UNSW), Sydney, Australia. Mahdi's research outcomes, which lie at the intersection of Internet-based computing technologies, can be in the form of conceptual models, system development methodologies, and decision-making frameworks. Before joining the academia, Mahdi has worked as a software engineer in different industry sectors including accounting, insurance, and publishing.

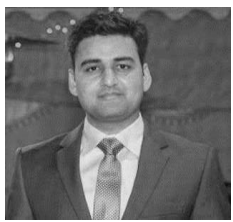

**Aakah Ahmad** is working as an Assistant Professor of Software Engineering at the College of Computer Science and Engineering, University of Ha'il, Saudi Arabia. Aakash's research interestes are in the area of software engineering and software architecture for mobile and pervasive systems.

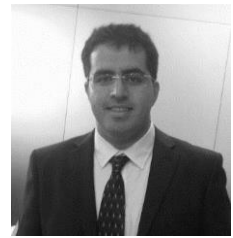

**Ali Behnaz** is Analytics Manager in the Advanced Analytics team of Citibank, Australia. He also holds an adjunct lecturer position with School of Computer Science and Engineering (CSE) at the University of New South Wales. His research output has manifested itself in the form of frameworks for enhancing implementation of machine learning use cases in organisations in the areas of finance, IoT, digital marketing and property valuation.

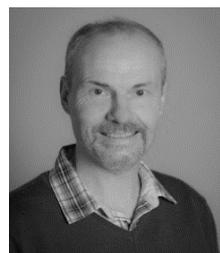

**John Grundy** is Australian Laureate Fellow and Professor of Software Engineering at Monash University, Australia. He has published widely in automated software engineering, domain-specific visual languages, model-driven engineering, software architecture, and empirical software engineering, among many other areas. He is Fellow of Automated Software Engineering and Fellow of Engineers Australia.

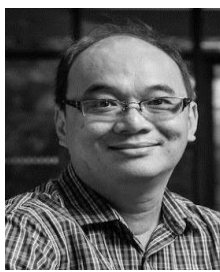

Willy Susilo received the Ph.D. degree in computer science from the University of Wollongong, Australia. He is currently a Professor and the Head of the School of Computing and Infor-mation Technology and the Director of the Institute of Cyberse-curity and Cryptology (iC2), University of Wollongong.